\documentclass[prd,showpacs,twocolumn,superscriptaddress,nofootinbib]{revtex4-2}
\usepackage{mathrsfs}
\usepackage{yfonts}
\usepackage{tipa}
\usepackage{bm}
\usepackage{bbm}
\usepackage{amsmath}
\usepackage{amssymb}
\usepackage{amsthm}
\usepackage{amsfonts}
\usepackage{mathtools}
\usepackage{latexsym}
\usepackage{relsize}
\usepackage[english]{babel}
\usepackage{booktabs}
\usepackage[utf8]{inputenc}
\usepackage{tikz}
\usepackage{array}
\usepackage{hyperref, color}
\usepackage{tikz}
\usetikzlibrary{shapes.geometric,arrows.meta,positioning,shapes.symbols}

\newcommand{\lp}{\left(}
\newcommand{\rp}{\right)}
\newcommand{\lb}{\left[}
\newcommand{\rb}{\right]}
\newcommand{\lag}{\left\langle}
\newcommand{\rag}{\right\rangle}
\newcommand{\ld}{\left.}
\newcommand{\rd}{\right.}

\newcommand{\bea}{\begin{eqnarray}}
\newcommand{\eea}{\end{eqnarray}}
\newcommand{\be}{\begin{equation}}
\newcommand{\ee}{\end{equation}}
\newcommand{\ba}{\begin{align}}
\newcommand{\ea}{\end{align}}

\newcommand{\D}{{\rm d}}%%%
\newcommand{\om}{\omega}

\newcommand{\nn}{\nonumber}

\newcommand{\bt}{\beta}

\newcommand{\vi}{\varphi}
\newcommand{\da}{\delta}

\newcommand{\en}{\epsilon}

\newcommand{\Sa}{\Sigma}

\newcommand{\apx}{\approx}

\begin{document}

\title{Gravitational radiation from binary systems with time varying masses}

\author{Teodora M. Matei}
	\email{teodora.maria.matei@stud.ubbcluj.ro}
	\affiliation{Astronomical Institute of Romanian Academy, Cluj-Napoca Branch, 19 Cire\c silor Street, 400487 Cluj-Napoca, Romania,}
	\author{Cristian A. Croitoru}
	\email{croitoru.lu.cristian@student.utcluj.ro}
	\affiliation{Department of Computer Science, Technical University of Cluj-Napoca, G. Baritiu Street 26-28, Cluj-Napoca 400027, Romania,}
\author{Tiberiu Harko,*}
\email[Corresponding author:]{tiberiu.harko@aira.astro.ro}
\affiliation{Astronomical Institute of Romanian Academy, Cluj-Napoca Branch, 19 Cire\c silor Street, 400487 Cluj-Napoca, Romania,}
\affiliation{Faculty of Physics, Babe\c s-Bolyai University, 1 Kog\u alniceanu Street,
	Cluj-Napoca, 400084, Romania}

%\affiliation{Corresponding author: Tiberiu Harko tiberiu.harko@aira.astro.ro}

\begin{abstract}
We consider the properties of the gravitational radiation emitted by  massive stars in binary systems with time varying mass, orbiting around each other in Keplerian orbits under the influence of the gravitational force. When this effect is combined to the quadrupole expression of the gravitational radiation, some supplementary terms in the standard formulae describing radiative effects do appear. By using the generalized quadrupole formalism, with time dependent masses, we obtain the expressions describing the time variation of the angular momentum, semimajor axis, eccentricity and periastron shift of the binary system. As a simple application of the developed formalism we consider in detail two cases of the mass variation, by assuming a linear and an exponential time dependence, respectively. The gravitational radiation characteristics corresponding to these models are fully investigated numerically with the help of a dedicated software package \href{https://github.com/croi900/GRAV-T}{GRAV-T}, specifically developed for the study of the gravitational radiation of the mass varying systems. The impact of the gravitational wave emission on the merger of two magnetars losing mass due to neutrino heated winds is also investigated in detail.  The obtained results could be important for the understanding of general relativistic effects in the case of the variation of the gravitational mass with time, which sensitively influences the properties of the gravitational radiation emission. 
\end{abstract}

%\pacs{04.50.Kd, 04.40.Dg, 04.20.Cv, 95.30.Sf}
\maketitle
%\tableofcontents

\section{Introduction}

Since the direct detection of the first gravitational wave signal GW150914 in September 2015 by the Laser Interferometer Gravitational wave Observatory (LIGO) \cite{Abbott, Abbott1}, the study of the gravitational radiation has become one of the fundamental fields of investigation in physics and astrophysics. Gravitational waves are free gravitational fields that are independent of bodies. Their existence was predicted by Poincar\'e \cite{Poin}, on the basis of special relativity and Lorentz transformations,  and by Einstein in 1916 \cite{Ein},  the year after the publication of his general theory of relativity. 

However for a long time, the physical reality of gravitational waves was considered as questionable. On account of the coordinate freedom of general relativity, it was assumed that gravitational waves were just coordinate waves \cite{George}. But this problem was solved by Pirani \cite{Pirani}, who interpreted gravitational waves as Weyl tensor waves carrying away energy and momentum. See \cite{Jorge} for a history of the discovery of the gravitational waves. 

Therefore, gravitational radiation could lead, for example,  to the orbital decay in a binary system consisting of a pulsar and a companion star, which could serve as an indirect method of observation of the wave emission. Even though the amplitude of the gravitational waves is extremely small, in 1975  Hulse and Taylor did observe the energy loss due to the emission of gravitational radiation  by the binary pulsar system PSR B1913+16,  during the inspiral phase \cite{Hulse}. The observation have confirmed at a high precision level the predictions of general relativity, with the ratio of the observed to predicted rate of orbital decay given by $0.997\pm 0.002$ \cite{Hulse1}. This result was improved in \cite{Hulse2}, where it was found  that the ratio of the observed value compared to the predicted value is $0.9983 \pm 0.0016$, indicating the presence of a 0.16\% disparity between theory and observations.  

After the discovery of PSR B1913+16, many other binary pulsars have been detected. Their observations are important not only for testing general relativity, but also for constraining modified gravity theories and alternative mechanisms of gravitational radiation \cite{test1,test2,test3,test4,test5,test6,test7,test8,test9,test10,test11,test12,test13,test14,test15}. For example, cubic Galileon models were tested in \cite{test7}, while data from well-timed binary pulsars have also been used to place bounds on the graviton mass, $m_g\leq 2\times 10^{-28}$ eV/c$^2$ at the 95\% confidence level. X-ray binary pulsars were also pointed out as useful  for testing modified gravity theories such as $f(Q,T)$ gravity \cite{test13}.

The observations of gravitational waves provide a unique access to the properties of the space-time in the strong-field and high-velocity regime, and allow the study of the properties of compact objects, like, for example, the equation of the state of neutron stars, black holes, black hole mergers etc. For a review of the physical properties of the persistent (continuous) gravitational waves  emitted by neutron stars see \cite{Univ}.

Even though the direct detection of gravitational radiation is extremely difficult, advances in numerical relativity and detector sensitivity have enabled a large number of observations by LIGO and Virgo \cite{Ligo}. The event GW150914 confirmed the existence of binary stellar-mass black hole systems, and many additional compact-binary mergers have since been reported \cite{Abbott2,Abbott3}. These include the binary neutron star merger GW190425 and systems with components in the neutron-star mass gap \cite{Abbott4,Shap}, as well as very massive black-hole mergers such as GW231123. Together, these observations suggest a rich variety of compact-binary formation channels, including routes to intermediate-mass black holes.

The rate of the emission of the gravitational energy from a system of two point masses
was calculated, with the use of the quadrupole formalism for gravitational radiation in the classic papers \cite{Peters, Peters1} for the case of elliptic motion, while in \cite{Hansen} and \cite{Turner} the case of the hyperbolic motion was considered. The results obtained in \cite{Peters, Peters1} have provided a very good explanation of the observation that the orbital period $P_b$ of the pulsar PSR 1913+16 is regularly decreasing by the amount
$\left(dPb/dt\right)_{{\rm obs}} = \left(-2.40 \pm 0.04\right) \times 10^{-12}$  \cite{TW}. The theoretical
value obtained from the formula in \cite{Peters} gives $\left(dPb/dt\right)_{{\rm th}} = \left(-2.403\pm 0.001\right)\times 10^{-12}$. This theoretical value agrees within 0.125\% with the observed one, and thus provides a powerful evidence for the existence of gravitational radiation. For other investigations of the quadrupole formalism and its validity see \cite{V1,V2,V3,V4,V5,V6}.

However, most of the analysis of the quadrupole formula and of the orbital decay in binary systems has been done under the assumption that the mass of the two components is a constant. Even if this approximation works well in isolated system, in many astrophysical situations one experiences a significant amount of mass loss by the components, or mass transfer inside the system. 

A typical example are the so called cataclysmic variables, composed of a white dwarf as the primary star and a low-mass main-sequence star or brown dwarf as the secondary \cite{Rez}.  In such a system, a mass overflow is induced from the inner Lagrange point onto the primary degenerate white dwarf, which happens due to the loss of angular momentum. If the period of the cataclysmic binaries is
longer than about 2 hours, then the loss of angular momentum can be explained by the interaction between the magnetic field and the stellar wind. In the case
of short-period binaries, with short periods less than about 2 hours, the only possibility to explain the loss of angular momentum is via gravitational radiation \cite{Rez}.

The mass transfer in binary systems consisting of white dwarfs was considered in \cite{Pac1,Pac2}. More complicated configurations have also been investigated, such as the hierarchical four-body system studied in \cite{Seto}, composed of a pair of mass-transferring white dwarf binaries. In a simplified model around the synchronous state of the two inner orbital periods, the system settles into a limit cycle with a small period gap, which produces a beat-like amplitude modulation of the emitted gravitational waves. Depending on the model parameters, the beat period can be as large as 1-10 years, making such variations potentially observable by space-based detectors.

Mass variation may also occur in binary systems consisting of two neutron stars. For example, a phase-transition-induced collapse can transform a neutron star into a hybrid star containing hadronic and strange quark matter \cite{Cheng}. Such a collapse may trigger large-amplitude stellar oscillations, producing pulsating neutrino and antineutrino fluxes. Because the neutrino energy and density at the peaks are much higher than in the non-oscillating case, the electron-positron pair creation rate increases strongly, which can eject surface layers that may later be accelerated to relativistic speeds.

The implications of the gravitational mass variation in binary system on the gravitational wave emission and orbital decay were first considered in \cite{Cheng1}, were the gravitational radiation from two time variable mass stars was considered. The total rates of the variation of the energy, angular momentum, semimajor axis, eccentricity and orbital period were obtained. The obtained results were applied to the case of the variation of the gravitational mass due to the spinning down as a result of the emission of electromagnetic radiation, which sensitively depends on the equations of state of the dense matter. The binary systems PSR 1913+16 and PSR 1534+12 were analyzed in detail, and, for different equations of state of nuclear matter, the corrections to the orbital decay due to gravitational radiation and to the spinning down of the pulsars were calculated. The obtained results did show that even in slowly rotating binary systems an improvement in the observational techniques could lead to the detection of the specific general relativistic effect of mass variation of pulsars due to spinning down through the study of orbital decay.

The additional contributions to the gravitational radiation from close binaries that arise from time-varying masses along with those produced by orbital motion were investigated in \cite{Holgado1}. Phase-dependent relations for these effects in the quadrupolar limit for binary point masses were derived, which reduce to the standard results when the mass of each component is taken to be constant. Gravitational radiation as a result of the mass variation can be orders of magnitude greater than that of the orbital motion.

In \cite{Holgado2} it was shown that gravitational waves  may be produced not only from the core-collapse supernova process, but also from the supernova mass loss and supernova natal kick during the transition from the pre-supernova to the post-supernova binary phase. The dynamical evolution of a binary at the time of the second supernova explosion was modeled with an equation of motion that accounts for the finite timescales of the supernova mass loss and the supernova natal kick. From the analysis of the dynamical evolution of the binary the gravitational burst signals associated with the supernova natal kicks were obtained. The gravitational energy emitted due to the supernova mass loss and natal kick may be a significant fraction, around 10\%, of the post-supernova binary's orbital energy, while for unbound post-supernova binaries, the gravitational energy radiated away is higher than that of bound binaries.

The coalescence (merger) of the two components in a binary system is an important source of gravitational radiation. After the formation of the binary system, the orbital separation of the components decreases gradually due to the long-term gravitational radiation emission, leading finally to the merger of the stars, and to a powerful burst of gravitational radiation. For the study of the various astrophysical aspects of the coalescence of the compact objects see \cite{C1,C2,Blanchet, C3,Nyadzani, C4,C5,C6,C7,C8,C9,C10,C11,C12,C13,C14}. 

It is the main goal of the present paper to extend and generalize the studies initiated in \cite{Cheng1, Holgado1,Holgado2} by taking into account in a systematic way the effects of the mass variation of the components in a binary system on the emission of the gravitational radiation. This leads to a generalization of the classic results of \cite{Peters} within the framework of a variable mass system. Our approach is also based on the quadrupole formalism, and on the assumption of the validity of the Newtonian approximation for the description of the orbital parameters.

 By assuming that both components of the binary system have a time dependent mass, we obtain first the variation of the energy of varying mass binary system due to gravitational radiation emission, as well as the averaged energy loss rate. The other important orbital parameters - the angular momentum loss rate, the variation of the semi-major axis of the orbit, as well as the variation of the period are also obtained in a general form. 
 
 As an astrophysical application of the considered results we pose the problem of the coalescence time in binary systems due to gravitational wave emission. To simplify the mathematical formalism we consider the coalescence problem for two specific mass variation laws, corresponding to a linear time variation and an exponential variation of the mass, respectively. As a concrete application of the obtained results we consider an astrophysically relevant system, consisting of two magnetars that lose mass through a neutrino-heated wind. 
We analyze the evolution of this system through the simulator \href{https://github.com/croi900/GRAV-T}{GRAV-T}, which follows the inspiral phase up to coalescence. Since our model treats the binary components as point masses and does not include hydrodynamic effects, it remains valid up to contact but cannot capture tidal deformation, mass transfer, or post-merger dynamics.
 
 Our results show the existence of a  difference in the properties of the gravitational radiation when the mass of the binary system varies, and can be considered as a function of time, as compared to the standard general relativistic results that assume a constant mass of the binary system.

 The present paper is organized as follows. We review the basics of the gravitational wave emission in standard general relativity in Section~\ref{sect1}. The gravitational wave emission from binary system with variable, time dependent mass, is considered in Section~\ref{variable_mass}. The coalescence time of the two components of the binary system in the presence of variable masses is obtained and analyzed in Section~\ref{sect3}. The specific case of a binary system consisting of two magnetars is investigated in Section~\ref{sect4}. We discuss and conclude our results in Section~\ref{sect5}.

\section{ Brief review of gravitational radiation emission from binary systems}\label{sect1}

In the present Section we briefly review the quadrupole description of gravitational radiation for constant point masses on Keplerian orbits, which provides the baseline for the variable-mass generalization developed below. We work in the weak-field, slow-motion limit of general relativity \cite{Landau, Straumann, Michele, Peters, Peters1}.

\subsection{Quadrupole radiation in linearized gravity}

Gravitational waves arise as solutions of the linearized Einstein equations in the weak-field, post-Minkowskian expansion \cite{Landau, Straumann, Michele}
\be
g_{\mu\nu}=\eta_{\mu\nu}+h_{\mu\nu}, \quad |h_{\mu\nu}|\ll 1,
\ee
where $\eta_{\mu\nu}={\rm diag}(1,-1,-1,-1)$. Infinitesimal coordinate changes $x^{\prime\mu}=x^{\mu}+\xi^{\mu}$ leave the linear theory form-invariant. In the Lorenz gauge the trace-reversed perturbation $\psi_{\mu\nu}$ satisfies the inhomogeneous wave equation
\be\label{waveeq}
\square\psi_{\mu\nu}=\frac{16\pi G}{c^4}T_{\mu\nu},
\ee
which reduces in vacuum to $\square\psi_{\mu\nu}=0$, with null wave vectors $k^\alpha k_\alpha=0$. The residual gauge freedom (subject to $\square\xi^\mu=0$) can be used to impose the transverse-traceless conditions
\be
h_{0\mu}=0,\quad h:=\eta^{\alpha\beta}h_{\alpha\beta}=0,\quad \partial^j h_{ij}=0,
\ee
leaving two independent polarization modes, $h_+$ and $h_\times$.

In the far zone the retarded solution of Eq.~(\ref{waveeq}) behaves as $1/R_0$. For non-relativistic sources, $v\ll c$, the dominant contribution comes from the mass quadrupole,
\begin{equation}
	M_{ij}(t) = \int \rho(t,\mathbf{x}')\,x'_i x'_j\,d^3x', \label{masstensor}
\end{equation}
\begin{equation}
	D_{ij} = 3M_{ij} - \delta_{ij}M_{kk},
\end{equation}
which is symmetric and traceless. The corresponding TT waveform is \cite{Blanchet}
\begin{equation}
	h_{ij}^{\rm TT} = \frac{2G}{c^4 R}\,\ddot{D}_{ij}\left(t - \frac{R}{c}\right) \Lambda_{ij,kl},
\end{equation}
with projector $\Lambda_{ij,kl}=P_{ik}P_{jl}-\frac12 P_{ij}P_{kl}$, $P_{ij}=\delta_{ij}-n_in_j$. For an observer along $\hat{z}$ the two polarizations reduce to
\begin{align}
	h_+ &= \frac{G}{c^4 R}\!\left(\ddot{D}_{xx} - \ddot{D}_{yy}\right), 
	\quad
	h_\times = \frac{2G}{c^4 R}\,\ddot{D}_{xy},
\end{align}
and the total strain is $h = \sqrt{h_+^2 + h_\times^2}$.

The effective energy density of a gravitational wave in the TT gauge is
\begin{equation}
	t^{00} = \frac{c^4}{32\pi G} \langle \dot{h}_{ij}^{\rm TT} \dot{h}_{ij}^{\rm TT} \rangle,
\end{equation}
where $\langle\cdot\rangle$ denotes an average over several wave periods. Integrating the associated flux over a sphere at radius $R$ and substituting the quadrupole waveform results in the Einstein quadrupole formula \cite{Landau, Straumann, Michele}
\begin{equation}
	-\frac{dE}{dt} = \frac{G}{45c^5}\langle\dddot{D}_{ij}\dddot{D}_{ij}\rangle.
\end{equation}
The associated angular-momentum loss rate is \cite{Landau}
\be
-\frac{\D L^i}{\D t}=\frac{2G}{45c^5}\en^{ijk}\ddot{D}_{ij}\dddot{D}_{kl}. \label{Lloss}
\ee
A short derivation of Eq.~(\ref{Lloss}) from the quadrupole power is collected in Appendix~\ref{AppLloss}.
\color{black}

\subsection{Gravitational radiation from binary systems} \label{binaries}

We consider a gravitationally bound point-mass binary with component masses $M_1$ and $M_2$, with the coordinates $(r_1\cos\vi, r_1\sin\vi)$ and $(-r_2\cos\vi, -r_2\sin\vi)$ in the $xy$ plane \cite{Peters, Peters1, Hansen}. The origin will be taken to be the center of the mass. We denote the distance between the two point-masses by $r$, the total mass by $M=M_1+M_2$ and the reduced mass as $\mu=M_1M_2/M$. Hence the position vectors of the stars are  $r_1=M_2r/M$ and $r_2=M_1r/M$, respectively. 

From the Newtonian analysis of the two-body problem in classical mechanics, we have \cite{Landau1}
\begin{gather}
r=\frac{a(1-e^2)}{1+e\cos\vi},\;\;
\dot{\vi}=\frac{1}{r^2}\sqrt{GMa\lp1-e^2\rp}\\
P_b=\int^{2\pi}_0\frac{\D\vi}{\dot{\vi}}=\frac{2\pi a^{3/2}}{\sqrt{GM}},\;\;
a=-\frac{GM_1M_2}{2E}, \label{period}\\
e^2=1+\frac{2EL^2 M}{G^2M_1^3M_2^3},
\end{gather}
where $P_b$ is the period of an orbit, $a$ is the semi-major axis, and $e$ is the eccentricity. In the $xy$ plane of spherical coordinate frame, we have $x^1=x=r\cos\vi$, $x^2=y=r\sin\vi$, and $x^3=z=0$, and
\be
\dot{D}_{ij}=\frac{\D D_{ij}}{\D\vi}\dot{\vi}.
\ee

\paragraph{Energy and angular-momentum losses.} The instantaneous quadrupole fluxes on the Keplerian orbit are \cite{Peters, Peters1}
\begin{align}
-\frac{\D E}{\D t}
=&\frac{8G^4 M_1^2 M_2^2 M}{15c^5a^5\lp 1-e^2\rp^5}\lp 1+e\cos\vi\rp^4\times\nn\\
&\lb 12\lp 1+e\cos\vi\rp^2+e^2\sin^2\vi\rb, \label{Eloss}
\end{align}
\begin{align}
-\frac{\D L}{\D t}&=-\frac{\D L_z}{\D t}
=\frac{8G^3M_1^2M_2^2}{5c^5a^3\left(1-e^2\right)^3}\sqrt{\frac{GM}{a\left(1-e^2\right)}}(1+e\cos\vi)^3\nn\\
&\times\lb2\lp1+e\cos\vi\rp\lp2+e\cos\vi\rp-e^2\sin^2\vi\rb. \label{angloss}
\end{align}
The corresponding orbit-averaged rates are
\begin{align}
-\left\langle\frac{\D E}{\D t}\right\rangle
=&\frac{32G^4 M_1^2 M_2^2 M}{5c^5a^5\lp1-e^2\rp^{7/2}}\lp 1+\frac{73}{24}e^2+\frac{37}{96}e^4\rp,
\end{align}
\begin{align}
&-\left\langle\frac{\D L}{\D t}\right\rangle
=\frac{32G^3M_1^2M_2^2}{5c^5a^3\lp1-e^2\rp^2}\sqrt{\frac{GM}{a}}\lp1+\frac{7}{8}e^2\rp.
\end{align}

\paragraph{Orbital decay due to gravitational radiation.} From Eq.~(\ref{period}) we can obtain the energy of the two body system as
\be
E=-\frac{GM_1M_2}{2a}, \label{avgE}
\ee
and the angular momentum expression
\be
L^2=\frac{GM_1^2M_2^2a(1-e^2)}{M},
\ee
the orbit-averaged evolution of the semi-major axis, eccentricity and period reads \cite{Peters,Peters1}
\begin{align}
\left\langle\frac{\D a}{\D t}\right\rangle
=&-\frac{64G^3M_1M_2M}{5c^5a^3\lp1-e^2\rp^{7/2}}\lp1+\frac{73}{24}e^2+\frac{37}{96}e^4\rp, \label{avg_a}
\end{align}
\be \label{avg_ecc}
\left\langle\frac{\D e}{\D t}\right\rangle=-\frac{304G^3M_1M_2Me}{15c^5a^4\lp1-e^2\rp^{5/2}}\lp1+\frac{121}{304}e^2\rp,
\ee
\begin{align}
\lag\frac{\D P_b}{\D t}\rag=&-\frac{192\pi G^2M_1M_2}{5c^5a^2\lp1-e^2\rp^3}\nn\\
&\times \sqrt{\frac{GM}{a\lp1-e^2\rp}}\lp1+\frac{73}{24}e^2+\frac{37}{96}e^4\rp,
\end{align}
or equivalently
\be
\lag\frac{\dot{P_b}}{P_b}\rag=-\frac{96G^3M_1M_2M}{5c^5a^4\lp1-e^2\rp^{7/2}}\lp1+\frac{73}{24}e^2+\frac{37}{96}e^4\rp.
\ee

The orbit averaging of Eqs.~(\ref{Eloss}) and (\ref{angloss}), and the derivation of Eqs.~(\ref{avg_a})--(\ref{avg_ecc}) from the Newtonian relations for $E$ and $L$, are given in Appendix~\ref{AppBinaries}.

\section{Gravitational radiation from binary systems with time variable gravitational mass} \label{variable_mass}

In the present Section we will present the basic formalism for the description of the effects of the mass variation on the gravitational radiation emission in binary systems. We will first estimate the effects on the quadrupole moment of the time varying mass, and we will obtain the basic corrections to these quantities. Then we will  derive the expressions of the gravitational power emission, of the angular momentum loss, and of the orbital decay in binary systems in the presence of time varying gravitational masses.

It should be emphasized that the masses vary without input or output of orbital energy and angular momentum, otherwise the energy change of the system is caused by more than gravitational radiation. In other words, we treat the mass variations as effectively instantaneous, while conserving the total orbital energy and angular momentum. We assume the binary masses change so slowly that $P_b\dot{M}_1\ll M_{c1}, P_b\dot{M}_2\ll M_{c2}$, or equivalently $P_b\dot{f}_1\ll 1, P_b\dot{f}_2\ll 1$. Under these conditions, the effective one body model continues to apply, so all analyses and results in this section remain valid.

\subsection{The two body problem for variable mass systems}

The Newtonian two body problem with variable mass has attracted a lot of interest for both its theoretical structure and its astrophysical applications \cite{T1,T2,T3,T4,T5}. For two point masses interacting through Newtonian gravity the orbits remain planar \cite{T3}, and one may take $z=\dot{z}=0$ without loss of generality. The specific angular momentum $h=x\dot{y}-\dot{x}y=r^{2}\dot{\theta}$ is conserved for isotropic mass loss, and the radial equation of the relative motion reads
\begin{equation}\label{eqmot}
\frac{d^{2}r}{dt^{2}}=-\frac{Gm}{r^{2}}+\frac{h^{2}}{r^{3}},
\end{equation}
with $m=m_1+m_2$. The orbital energy per unit mass,
\begin{equation}
E=\frac{1}{2}\dot{r}^{2}+\frac{1}{2}\frac{h^{2}}{r^{2}}-\frac{Gm}{r},
\end{equation}
then varies only because of the mass change \cite{T3},
\begin{equation}
\frac{dE}{dt}=-G\frac{\dot{m}}{r}.
\end{equation}
In terms of $u=1/r$ this dynamics can be rewritten as a generalized Binet equation,
\begin{equation}
\frac{d^{2}u}{d\theta ^{2}}+u=\frac{Gm}{h^{2}},
\end{equation}
whose $\theta$-derivative encodes the mass-loss rate,
\begin{equation}
\frac{d}{d\theta }\left( \frac{d^{2}u}{d\theta ^{2}}+u\right) =\frac{G}{h^{3}}\frac{\dot{m}}{u^{2}}.
\end{equation}
For the time variation of the mass one could assume a relation of the form $\dot{m}=-k(m-m_r)^n$, with the latter equation being solvable only perturbatively. For the orbital elements of the osculating orbital see \cite{T1,T2,T3}.

A complementary question is the secular evolution of the separation itself in time-varying binary systems. For isotropic mass variation, $\ddot{\vec{r}}=-GM(t)\vec{r}/r^3$ implies conservation of $J=\mu\sqrt{GM(t)r}$. In the equal-mass case this gives \cite{T4}
\begin{equation}
\frac{dr}{dt}=-3\frac{\dot{m}}{m}r,
\end{equation}
so that mass loss ($\dot{m}<0$) drives the companions apart. The power associated to mass variation is
\begin{equation}
P_{m}=\frac{5}{2}\frac{G\dot{m}m}{r},
\end{equation}
while gravitational radiation removes energy at the quadrupole rate \cite{Michele}
\begin{equation}
P_{gw}=\frac{64}{5}\frac{G^{4}}{c^{5}}\frac{m^{5}(t)}{r^{5}}.
\end{equation}
Energy balance, $-\dot{E}=P_m+P_{gw}$, then gives the combined evolution equation \cite{T4}
\begin{equation}
\frac{dr}{dt}=-\frac{128}{5}\frac{G^{3}}{c^{5}}\frac{m^{3}(t)}{r^{3}}-3\frac{\dot{m}}{m}r,
\end{equation}
which describes a mass-loss dominated regime at large separation and a radiation dominated inspiral at small $r$.

Finally, because the Lagrangian depends explicitly on time through $m(t)$, energy is not a Noether charge. For isotropic mass loss the center of mass remains at rest and rotational/spatial symmetries may survive. For detailed analysis on Noether's theorem of rotational relativistic variable-mass systems, including inverse theorems and explicitly time-dependent conserved quantities, see \cite{T5}.
\color{black}

\subsection{Quadrupole moments with time varying masses}

 We add the index $c$ in the equations in Section~\ref{variable_mass}, to distinguish the case of stars with constant mass from the case of stars with time varying gravitational mass. We assume that \cite{Cheng}
\begin{gather}
M_1(t)=M_{c1}f_1(t),
M_2(t)=M_{c2}f_2(t),\\
M(t)=M_{c}f_M(t),
\mu(t)=\mu_cf_\mu(t).
\end{gather}
Obviously,
\begin{gather}
f_M(t)=\frac{M_{c1}}{M_{c}}f_1(t)+\frac{M_{c2}}{M_{c}}f_2(t),\\
f_\mu(t)=\frac{M_cf_1(t)f_2(t)}{M_{1c}f_1(t)+M_{2c}f_2(t)}.
\end{gather}

The quadrupole moment of the variable-mass binary is $D_{ij}=D_{cij}f_\mu$, so that its time derivatives acquire additional terms proportional to $\dot{f}_\mu$, $\ddot{f}_\mu$ and $\dddot{f}_\mu$. Relating the constant-mass derivatives evaluated along the instantaneous Keplerian orbit to the reference (constant-mass) derivatives $(D_{ij})_c,\ldots,(\dddot{D}_{ij})_c$ yields the linear transformation
\be
\begin{pmatrix}
D_{cij}\\
\dot{D}_{cij}\\
\ddot{D}_{cij}\\
\dddot{D}_{cij}
\end{pmatrix}
=
\begin{pmatrix}
1&0&0&0\\
0&\sqrt{f_M}&0&0\\
0&\frac{\dot{f}_M}{2\sqrt{f_M}}&f_M&0\\
0&\frac{2\ddot{f}_Mf_M-\dot{f}_M^2}{4f_M^{3/2}}&\frac{3\dot{f}_M}{2}&f_M^{3/2}
\end{pmatrix}
\begin{pmatrix}
(D_{ij})_c\\
(\dot{D}_{ij})_c\\
(\ddot{D}_{ij})_c\\
(\dddot{D}_{ij})_c
\end{pmatrix},
\ee
which we denote by ${\bf A}$. Combining this map with the chain rule for $\dddot{D}_{ij}$ and $\ddot{D}_{ij}$ gives
\begin{align}
\dddot{D}_{ij}&={\bf F}\,{\bf D}'_c,
\qquad
\ddot{D}_{ij}={\bf F}'\,{\bf D}'_c,
\end{align}
where ${\bf D}'_c=((D_{ij})_c,(\dot{D}_{ij})_c,(\ddot{D}_{ij})_c,(\dddot{D}_{ij})_c)^{\rm T}$ and
\bea
\hspace{-0.7cm}F_1&=&\dddot{f}_\mu,\\
\hspace{-0.7cm}F_2&=&3\ddot{f}_\mu\sqrt{f_M}+3\frac{\dot{f}_\mu\dot{f}_M}{2\sqrt{f_M}}+\frac{f_\mu }{4f_M^{3/2}}(2\ddot{f}_Mf_M-\dot{f}_M^2),\quad\\
\hspace{-0.7cm}F_3&=&3\dot{f}_\mu f_M+3\frac{\dot{f}_Mf_\mu}{2},\\
\hspace{-0.7cm}F_4&=&f_M^{3/2}f_\mu , \\
F'_1&=&\ddot{f}_\mu, \;
F'_2=2\dot{f}_\mu\sqrt{f_M}+\frac{f_\mu\dot{f}_M}{2\sqrt{f_M}}, \;
F'_3=f_Mf_\mu.
\eea
Consequently,
\begin{align}
&\ddot{D}_{ij}\dddot{D}_{kl}=\lb(\ddot{D}_{ij})_cF'_3+2(\dot{D}_{ij})_cF'_2+(D_{ij})_cF'_1\rb\times\nn\\
&\lb(\dddot{D}_{kl})_cF_4+3(\ddot{D}_{kl})_cF_3+3(\dot{D}_{kl})_cF_2+(D_{kl})_cF_1\rb, \label{d3Dsquared}
\end{align}
and the squared third derivative $\dddot{D}_{ij}^2$ follows by the same expansion. The intermediate algebra leading to these relations is given in Appendix~\ref{AppQuad}.

\subsection{Gravitational radiation power for time variable gravitational masses}

From Eq.~(\ref{Eloss}) we obtain for the gravitational energy radiated by a binary system with time varying masses the expression
\begin{align}
-\frac{\D E}{\D t}=&\frac{G}{45c^5}\lp \dddot{D}_{xx}^2+2\dddot{D}_{xy}^2+\dddot{D}_{yy}^2+\dddot{D}_{zz}^2\rp\nn\\
=&-\lp\frac{\D E}{\D t}\rp_cF_4^2+\frac{G}{45c^5}\lp 6A_1F_3 F_4\rd\nn\\
&+6A_2F_2F_4+9A_3F_3^2+2A_4F_1 F_4\nn\\
&+18A_5F_2F_3+6A_6F_1F_3+9A_7F_2^2\nn\\
&\ld+6A_8F_1F_2+A_9F_1^2\rp,
\end{align}
\iffalse\begin{align}
-\frac{\D E}{\D t}=&\frac{G}{45c^5}\lp \dddot{D}_{xx}^2+2\dddot{D}_{xy}^2+\dddot{D}_{yy}^2+\dddot{D}_{zz}^2\rp\nn\\
=&-\lp\frac{\D E}{\D t}\rp_cf_\mu^2+\frac{2G}{15c^5}A_1f_\mu\dot{f}_{\mu}+\frac{2G}{15c^5}A_2f_\mu\ddot{f}_{\mu}\nn\\
&+\frac{G}{5c^5}A_3\dot{f}_{\mu}^2+\frac{2G}{45c^5}A_4f_\mu\dddot{f}_{\mu}+\frac{2G}{5c^5}A_5\dot{f}_\mu\ddot{f}_{\mu}\nn\\
&+\frac{2G}{15c^5}A_6\dot{f}_\mu\dddot{f}_{\mu}+\frac{G}{5c^5}A_7\ddot{f}_{\mu}^2+\frac{2G}{15c^5}A_8\ddot{f}_\mu\dddot{f}_{\mu}\nn\\
&+\frac{G}{45c^5}A_9\dddot{f}_{\mu}^2\nn\\
\end{align}\fi
where the constant mass part is expressed as
\begin{align}
-\lp\frac{\D E}{\D t}\rp_c=&\frac{8G^4 M_{1c}^2 M_{2c}^2 M_c}{15c^5a^5\lp 1-e^2\rp^5}\lp 1+e\cos\vi\rp^4\times\nn\\
&\lb 12\lp 1+e\cos\vi\rp^2+e^2\sin^2\vi\rb.
\end{align}

In order to calculate the terms $A_i$, $i=1,2,...,9$ we use the reduced quadrupole moments, whose components are obtained from $D_{ij} = 3M_{ij} - \delta_{ij} M_{kk}$ by considering the same binary system as in Section~\ref{binaries}, with
\bea\label{Dxx}
&&D_{xx}=\mu r^2(3\cos^2\vi-1), \quad D_{yy}=\mu r^2(3\sin^2\vi-1) \nn \\
&&D_{xy}=\mu r^2(3\sin\vi\cos\vi), \quad D_{zz}=-\mu r^2,
\eea
The explicit expressions of the coefficients $A_i$, $i=1,...,9$ are presented in Appendix~\ref{DefA}. 

Since  we assume that the masses of the binary stars vary very slowly, we neglect their variation in a single orbital period. As a consequence  of this assumption we can regard $F_1\sim F_4$ as constants in the integrations below.

The average energy loss rate through gravitational energy emission is obtained as
\begin{align}\label{eqloss}
-\lag\frac{\D E}{\D t}\rag&=\frac{1}{P_b}\int^{2\pi}_0\frac{\D E}{\D t}\frac{1}{\dot{\vi}}\D\vi\nn\\
=&\frac{\sqrt{GM}}{2\pi a^{3/2}}\int^{2\pi}_0\frac{\D E}{\D t}\sqrt{\frac{a\lp1-e^2\rp}{GM}}\frac{a\lp1-e^2\rp}{\lp1+e\cos\vi\rp^2}\D\vi\nn\\
=&\frac{(1-e^2)^{3/2}}{2\pi}\int^{2\pi}_0\frac{\D E}{\D t}\frac{\D \vi}{(1+e\cos\vi)^2}.
%=&-F_4^2\lag\frac{\D E}{\D t}\rag_c+\frac{G(1-e^2)^{3/2}}{90\pi c^5}\lb 6F_3F_4\int^{2\pi}_0\frac{A_1\D \vi}{(1+e\cos\vi)^2}\rd\nn\\
%&+6F_2F_4\int^{2\pi}_0\frac{A_2\D \vi}{(1+e\cos\vi)^2}+9F_3^2\int^{2\pi}_0\frac{A_3\D \vi}{(1+e\cos\vi)^2}\nn\\
%&+2F_1 F_4\int^{2\pi}_0\frac{A_4\D \vi}{(1+e\cos\vi)^2}+18F_2F_3\int^{2\pi}_0\frac{A_5\D \vi}{(1+e\cos\vi)^2}\nn\\
%&+6F_1F_3\int^{2\pi}_0\frac{A_6\D \vi}{(1+e\cos\vi)^2}+9F_2^2\int^{2\pi}_0\frac{A_7\D \vi}{(1+e\cos\vi)^2}\nn\\
%&+6F_1F_2\int^{2\pi}_0\frac{A_8\D \vi}{(1+e\cos\vi)^2}\ld+F_1^2\int^{2\pi}_0\frac{A_9\D \vi}{(1+e\cos\vi)^2}\rb.
\end{align}

The results of the integrations in the expression above are presented in  Appendix~\ref{app1}. Therefore we obtain for the average gravitational power emitted by a binary system with time varying masses the expression
\begin{align}
-\lag\frac{\D E}{\D t}\rag%=-F_4^2\lag\frac{\D E}{\D t}\rag_c+\frac{G(1-e^2)^{3/2}}{90\pi c^5}\nn\\
%&\lb\frac{48\pi G^2M_c^2\mu_c^2\lp4-\sqrt{1-e^2}\rp}{a^2(1-e^2)^2}\lp9F_3^2-6F_2F_4\rp\rd\nn\\
%&+\frac{12\pi GM_c\mu_c^2a\lp3-e^2\rp}{(1-e^2)^{3/2}}\lp9F_2^2-6F_1F_3\rp\nn\\
%&\ld+\frac{3\pi\mu_c^2a^4\lp8+40e^2+15e^4\rp}{2(1-e^2)^{3/2}}F_1^2\rb\nn\\
&=-F_4^2\lag\frac{\D E}{\D t}\rag_c+\frac{G\mu_c^2}{10c^5}\nn\\
&\times \lb\frac{16G^2M_c^2\lp4-\sqrt{1-e^2}\rp}{a^2\sqrt{1-e^2}}\lp3F_3^2-2F_2F_4\rp\rd\nn\\
&+4GM_ca\lp3-e^2\rp\lp3F_2^2-2F_1F_3\rp\nn\\
&\ld+\frac{1}{6}a^4\lp8+40e^2+15e^4\rp F_1^2\rb. \label{dE_mass}
\end{align}

\subsection{Angular momentum variation rate for time variable gravitational masses}

Similarly, from Eq.~(\ref{angloss}) we obtain the average angular momentum loss rate in the time varying mass system as
\begin{align}
-\frac{\D L}{\D t}&=\frac{2G}{45c^5}\lb\ddot{D}_{xy}\lp\dddot{D}_{yy}-\dddot{D}_{xx}\rp-\dddot{D}_{xy}\lp\ddot{D}_{yy}-\ddot{D}_{xx}\rp\rb\nn\\
=&\frac{2G}{45c^5}\lb\lp\ddot{D}_{xy}\dddot{D}_{yy}-\dddot{D}_{xy}\ddot{D}_{yy}\rp-\lp\ddot{D}_{xy}\dddot{D}_{xx}-\dddot{D}_{xy}\ddot{D}_{xx}\rp\rb.
\end{align}

The expressions of the terms containing the products of the time derivatives of the quadrupole moments are presented in Appendix~\ref{AppQuad}.
 Hence
\begin{align}
-\frac{\D L}{\D t}=&-F'_3 F_4\lp\frac{\D L_z}{\D t}\rp_c+\frac{2G}{45c^5}\lb F'_2F_4B_1+F'_1F_4B_2\rd\nn\\
&+\lp F'_3F_2-2F'_2F_3\rp B_3+\lp2F'_2F_1-3F'_1F_2\rp B_4\nn\\
&\ld+\lp F'_3F_1-3F'_1F_3\rp B_5\rb,
\end{align}
where the coefficients $B_i$, $i=1,\ldots,5$, are defined and calculated in Appendix~\ref{DefB}.

The average of the angular momentum loss via the gravitational radiation is defined according to
\bea\label{eqangloss}
-\lag\frac{\D L_z}{\D t}\rag&=&\frac{1}{P_b}\int^{2\pi}_0\frac{\D L_z}{\D t}\frac{1}{\dot{\vi}}\D\vi \nonumber\\
&=&\frac{(1-e^2)^{3/2}}{2\pi}\int^{2\pi}_0\frac{\D L_z}{\D t}\frac{\D \vi}{(1+e\cos\vi)^2}.
%=&-F'_3 F_4\lag\frac{\D L_z}{\D t}\rag_c+\frac{G(1-e^2)^{3/2}}{45\pi c^5}\nn\\
%&\lb F'_2F_4\int^{2\pi}_0\frac{B_1\D \vi}{(1+e\cos\vi)^2}+F'_1F_4\int^{2\pi}_0\frac{B_2\D \vi}{(1+e\cos\vi)^2}\rd\nn\\
%&+\lp F'_3F_2-2F'_2F_3\rp\int^{2\pi}_0\frac{B_3\D \vi}{(1+e\cos\vi)^2}\nn\\
%&+\lp2F'_2F_1-3F'_1F_2\rp\int^{2\pi}_0\frac{B_4\D \vi}{(1+e\cos\vi)^2}\nn\\
%&\ld+\lp F'_3F_1-3F'_1F_3\rp\int^{2\pi}_0\frac{B_5\D \vi}{(1+e\cos\vi)^2}\rb.
\eea

The results of the integrations in the above relation are presented in Appendix~\ref{app2}.  Hence we find our final result in the form
\bea\label{dL_mass}
-\lag\frac{\D L}{\D t}\rag
&=&-F'_3 F_4\lag\frac{\D L_z}{\D t}\rag_c-\frac{G\mu_c^2}{5c^5}\sqrt{\frac{GM_c(1-e^2)}{a}}\nn\\
&&\times \left[ 8GM_c(F'_1F_4+3F'_3F_2-6F'_2F_3)\rd\nn\\
&&\ld+a^3(2+3e^2)\lp2F'_2F_1-3F'_1F_2\rp\right]. 
\eea

\subsection{Decay of the orbital parameters in a binary system with time varying masses}

The average of the time derivative of $E=-GM_1M_2/2a$, obtained from Eq.~(\ref{avgE}), giving the total energy in a binary system is
\be
\left\langle\frac{\D E}{\D t}\right\rangle=\frac{GM_{c1}M_{c2}f_1f_2}{2a}\left[\frac{1}{a}\left\langle\frac{\D a}{\D t}\right\rangle-\lp\frac{\dot{f_1}}{f_1}+\frac{\dot{f_2}}{f_2}\rp\right].
\ee
\paragraph{Variation of the semi-major axis.} Hence, the variation rate of semi-major axis is given by
	\begin{align}
	\left\langle\frac{\D a}{\D t}\right\rangle=&\frac{2a^2}{GM_{c1}M_{c2}f_1f_2}\left\langle\frac{\D E}{\D t}\right\rangle+a\lp\frac{\dot{f_1}}{f_1}+\frac{\dot{f_2}}{f_2}\rp\nn\\
	%=&\frac{2a^2}{GM_{c1}M_{c2}f_1f_2}\lb F_4^2\lag\frac{\D E}{\D t}\rag_c-\frac{G\mu_c^2}{10c^5}\rd\nn\\
	%&\lp\frac{16G^2M_c^2\lp4-\sqrt{1-e^2}\rp}{a^2\sqrt{1-e^2}}\lp3F_3^2-2F_2F_4\rp\rd\nn\\
	%&+4GM_ca\lp3-e^2\rp\lp3F_2^2-2F_1F_3\rp\nn\\
	%&\ld\ld+\frac{1}{6}a^4\lp8+40e^2+15e^4\rp F_1^2\rp\rb
	%+a\lp\frac{\dot{f_1}}{f_1}+\frac{\dot{f_2}}{f_2}\rp\nn\\
	=&\frac{F_4^2}{f_1f_2}\lag\frac{\D a}{\D t}\rag_c-\lb\frac{\mu_c}{5c^5 f_1f_2}\rd\nn\\
	&\times \lp\frac{16G^2M_c\lp4-\sqrt{1-e^2}\rp}{\sqrt{1-e^2}}\lp3F_3^2-2F_2F_4\rp\rd\nn\\
	&+4G a^3\lp3-e^2\rp\lp3F_2^2-2F_1F_3\rp\nn\\
	&\ld\ld+\frac{a^6}{6M_c}\lp8+40e^2+15e^4\rp F_1^2\rp\rb
	+a\lp\frac{\dot{f_1}}{f_1}+\frac{\dot{f_2}}{f_2}\rp. \label{dadt}
	\end{align}

\paragraph{Averaged eccentricity evolution.} The average of Eq.~(\ref{avg_ecc}) with respect to time, after substitution of Eq.~(\ref{dE_mass}) and Eq.~(\ref{dL_mass}), is
\begin{align}
&2e\left\langle\frac{\D e}{\D t}\right\rangle=\frac{2M_c}{G^2M_{c1}^3M_{c2}^3}\lp\frac{f_M}{f_1^3f_2^3}L^2\left\langle\frac{\D E}{\D t}\right\rangle\rd\nn\\
&\ld+2EL\frac{f_M}{f_1^3f_2^3}\left\langle\frac{\D L}{\D t}\right\rangle+EL^2\frac{\dot{f}_M-3f_M\lp\frac{\dot{f}_1}{f_1}+\frac{\dot{f}_2}{f_2}\rp}{(f_1f_2)^3}\rp.
\end{align}

The expression of the average eccentricity for a binary system with time varying masses is presented in Appendix~\ref{ecc}. 

\paragraph{Variation of the orbital period.} From Eq.~(\ref{period}), we obtain the variation rate of the orbital period as
\begin{align}
\lp\frac{\D P_b}{\D t}\rp=&\frac{2\pi}{\sqrt{GM_c}}\frac{\D}{\D t}\lp\frac{a^{3/2}}{f_M^{1/2}}\rp\nn\\
=&\frac{2\pi}{\sqrt{GM_c}}\sqrt{\frac{a}{f_M}}\frac{3\dot{a}f_M-\dot{f}_Ma}{2f_M}.
\end{align}

Therefore, the time average of the variation rate of the period in the binary system with varying masses is
\begin{align}
\lag\frac{\D P_b}{\D t}\rag=&\frac{2\pi}{\sqrt{GM_c}}\sqrt{\frac{a}{f_M}}\lp\frac{3}{2}\lag\frac{\D a}{\D t}\rag-\frac{a\dot{f}_M}{2f_M}\rp. \label{P_rate}
\end{align}

Finally, we obtain for the variation rate of the orbital period
\be
\lag\frac{\dot{P_b}}{P_b}\rag=\frac{3}{2a}\lag\frac{\D a}{\D t}\rag-\frac{\dot{f_M}}{2f_M},
\ee
where $\lag\frac{\D a}{\D t}\rag$ can be substituted by Eq.~(\ref{dadt}).

\section{Astrophysical applications: the coalescence time in binary systems with time varying masses}\label{sect3}

In the present Section we will consider the important problem of the coalescence of two stars moving together in a binary system with time varying masses. After briefly reviewing the coalescence problem in standard general relativity, we will proceed to the investigation of the effects induced by the mass variation on the coalescence time in a binary system.

\subsection{Mass loss models in astrophysical systems}

Many types of stars have strong stellar winds or experience mass loss processes that determine their evolution, and even the black hole mass function \cite{N1,N2}. Mass loss also strongly influences the stellar envelope structure, which is fundamental for both single-star and binary evolution, and for the predictions of the gravitational wave emission. The mass-loss rate depends in general on $(M,R,L)$, the Eddington factor $\Gamma$, $T_{eff}$, and the composition $Z$. For simple models we retain only $(M,R,L)$ and take \cite{N1}
\be\label{MLR}
\frac{dM}{dt}=-KM^aL^bR^c,
\ee
where $(K,a,b,c)$ are constants. From a sample of 247 stars, \cite{N1} obtained
\bea\label{MLR1}
\frac{dM}{dt}=-9.63\times 10^{-15}&\times& \left(\frac{M}{M_\odot}\right)^{0.16}\times \left(\frac{L}{L_\odot}\right)^{1.42}\nonumber\\
&\times& \left(\frac{R}{R_\odot}\right)^{0.81}\;M_\odot/{\rm yr},
\eea 
which describes mass loss over the Hertzsprung--Russell diagram. Since $a$ is small in Eq.~(\ref{MLR1}), the limit $a\rightarrow 0$ with roughly constant luminosity yields $\dot{M}\approx -KL^bR^c$, and hence the linear model
\be
M(t)=M(0)-kt,
\ee  
with $k=L^bR^c$, appropriate for comparatively mild mass loss.

Extreme mass loss can also occur, such as in the case of the giant eruption of $\eta$ Car, which removed $\sim 10 M_\odot$ in about 20 years \cite{N2}. For massive stars one may take $a\rightarrow 1$ in Eq.~(\ref{MLR}), giving
\be
M(t)=M(0)\exp\left(-\int_0^t{L^b(t)R^c(t)dt}\right).
\ee
The mean-value theorem, $\int_0^t L^b R^c\,dt=\bigl(L^b R^c\bigr)_{t_c}t$ with $0<t_c<t$, then recovers the exponential model
\be
M(t)=M_0e^{-\omega t}.
\ee 

There are many mass loss processes that play an important role in astrophysics. In Section~\ref{sect4} we will investigate the impact on gravitational wave emission of the mass loss in binary magnetar systems due to neutrino-heated winds. A similar approach can be developed for the study of the impact of the URCA process \cite{Haen} on neutron stars in a binary system. If a neutron star contains a central core in which the direct URCA-process is active, then the cooling timescale of the star becomes shorter by many orders of magnitude. In pulsars and magnetars, astrophysical objects having strong magnetic fields, the  rotational kinetic energy of the star generates a relativistic wind of electrons and positrons. Due to the spin-down of  pulsars they continuously lose a small fraction of their mass in the form of particles and synchrotron radiation \cite{Smith, Cheng1}.  

Hence, mass loss processes play an important role in the dynamical evolution of binary systems \cite{P1,P2, P3, P4}.  There are several astrophysical effects contributing to mass loss for  pulsars, which are rapidly rotating and high density neutron stars. The strong magnetic field and the high rotational energy determines the emission of an intense  wind of relativistic particles (formed of ions and electrons), which leads to a decrease of the mass of the pulsar over time \cite{P1,P2,P3, P4}. A more important effect appears in pulsar-ordinary star binary system, where the  high-energy radiation of the pulsar and the pulsar winds interact with the companion star, removing its outer layers, and determining  extreme mass losses causing the destruction  of the companion star and the formation of a millisecond pulsar.

The mass loss rates in binary pulsar-companion star system are estimated to be of the order of $\dot{M}\sim 3.7-4.8 \times 10^{-7}M_\odot/{\rm yr}$ \cite{P1,P2}. This value was obtained for the $\gamma$-ray binary pulsar LS 5039 by using a coordinated space-based photometric and ground-based spectroscopic observing campaign \cite{P1, P2}. By assuming the linear mass model loss, this value would imply $\dot{M}/M_\odot\sim k\sim 1.17-1.52\times 10^{-14}\;{\rm s^{-1}}$. In the Black Widow binary system containing the millisecond pulsar B1957+20 the mass loss of the companion star is estimated to be around $\dot{M}\sim 10^{-10}M_\odot/{\rm yr}$ \cite{P3,P4}, giving for the case of the linear model $\dot{M}/M_\odot\sim k\sim 3\times 10^{-18}\;{\rm s^{-1}}$. Binary systems containing redback pulsars can have mass loss rates of the order of $10^{-11}-10^{-9}M_\odot/{\rm yr}$ \cite{P3,P4}.  

\color{black}

\subsection{The coalescence time in standard general relativity}

Since the radiation of gravitational waves is very small, their effects can't be observed in a short time. We are more interested in the effects with a long time accumulation. For the evolution of binary compact objects with time-independent mass, the energy carried away by gravitation waves during the inspiral will lead to the coalescence of this binary system. We calculate first the coalescence time in general relativity \cite{Nyadzani}.

From Eq.~(\ref{avg_a}) we obtain first,
\begin{align}
\left\langle\frac{\D a}{\D t}\right\rangle_c=-\frac{\bt}{a^3\lp1-e^2\rp^{7/2}}\lp1+\frac{73}{24}e^2+\frac{37}{96}e^4\rp, \label{a_coal}
\end{align}
while  Eq.~(\ref{avg_ecc}) gives
\be
\left\langle\frac{\D e}{\D t}\right\rangle_c=-\frac{19\bt e}{12a^4\lp1-e^2\rp^{5/2}}\lp1+\frac{121}{304}e^2\rp, \label{e_coal}
\ee
where we have denoted
\be
\bt=\frac{64G^3M_{1c}M_{2c}M_c}{5c^5}.
\ee

In the following, we omit the notation for averaging, as the previously mentioned quantities represent their time averaged values over a complete orbital period. The gravitational coalescence of the binary system occurs when the semi-major axis vanishes, $a=0$. The limiting case of circular orbits are of particular interest, characterized by vanishing eccentricity, $e=0$. After substituting this condition in Eq.~(\ref{avg_ecc}), the resulting equation admits an exact analytical solution for the coalescence timescale, which can be obtained from Eq.~(\ref{a_coal}) 
\begin{gather}
\frac{\D a}{\D t}=-\frac{\bt}{a^3}, \quad
\frac{\D a^4}{4}=-\beta dt,
\end{gather}
giving for the standard coalescence time the expression
\begin{gather}
T_c(a_0,0)=\int^{T_c}_0\D t=\int^0_{a_0^4}\frac{\D t}{\D a^4}\D a^4=\frac{a_0^4}{4\bt}. \label{T_coal}
\end{gather}

For eccentric binaries the relation $a(e)$ obtained from Eqs.~(\ref{a_coal},\ref{e_coal}) can be integrated to give an exact coalescence time in terms of Appell functions \cite{Nyadzani},
\be
T"'(a_0,e_0)=T_c\frac{(1-e_0^2)^{7/2}}{(1-e_0^{7/4})^{1/5}\lp1+\frac{121}{304}e_0^2\rp}.
\ee
\color{black}

\subsection{Coalescence time for time varying mass compact objects}

When considering time-varying masses in a binary system, the coalescence time can still be determined by accounting for mass evolution functions, $f_1(t)$ and $f_2(t)$. Therefore, by solving the coupled system of equations for the semi-major axis and eccentricity to determine $a$ as a function of $e$, and by substituting back into $\D e/\D t$ and integrating from the initial eccentricity $e_0$ down to 0, one can find the coalescence time for time varying systems.  

However, this approach involves increasingly complicated calculations as the mass variation functions are not explicitly specified, making a general solution hard to obtain in most astrophysical scenarios. 

To simplify the computation, we assume that both components evolve with the same time dependence, so that $f_1(t) = f_2(t) = f(t)$. This leads to the following relations,
\bea
f_{M}&=&f_{\mu }=f_{1} =f_{2}=f, \\
F_{1}&=&\dddot{f}, F_{2}=\sqrt{f}\lp\frac{7}{2}\ddot{f}+\frac{5}{4}\frac{%
\dot{f}^{2}}{f}\rp, \\
 F_{3}&=&\frac{9}{2}\dot{f}f,F_{4}=f^{2}\sqrt{f}, \\
F_{1}^{\prime } &=&\ddot{f},F_{2}^{\prime }=\frac{5}{2}\sqrt{f}\dot{f}%
,F_{3}^{\prime }=f^{2}. 
\eea

In the case of a circular orbit $(e_0=0)$, by substituting $e=0$ in Eq.~(\ref{dadt}) we obtain
\begin{align}\label{vara}
\frac{\D a}{\D t}
=&-\frac{64G^3M_cM_{c1}M_{c2}}{5c^5a^3}f^3+2a\frac{\dot{f}}{f}-\lb\frac{\mu_c}{5c^5}\rd\nn\\
&\times \lp48G^2M_c\lp\frac{233}{4}\dot{f}^2-7f\ddot{f}\rp+36Ga^3\rd\nn\\
&\times \lp\frac{49\ddot{f}^2}{4f}+\frac{25\dot{f}^4}{16f^3}+\frac{35\dot{f}^2\ddot{f}}{4f^2}-\frac{3\dddot{f}\dot{f}}{f}\rp
\ld\ld+\frac{4a^6}{3M_c}\frac{\dddot{f}^2}{f^2}\rp\rb.
\end{align}

\subsubsection{The linear time variation of the mass}

If $f(t)$ can be approximate by a linear function, then $\dot{f}$ is equal to a constant. From the definition of $f=M/M_c$, we know that $f(0)=1$. Using the linear approximation, we assume that the mass function is given by
\be
f(t)\apx 1+kt,
\ee
where $k$ is a small constant. Then Eq.~(\ref{vara}) giving $(\D a/\D t)$ in the presence of time varying masses becomes
\begin{align}\label{eqlin}
\frac{\D a}{\D t}
=&-\frac{64G^3M_cM_{1c}M_{2c}}{5c^5}\frac{(1+kt)^3}{a^3}+\frac{2ka}{1+kt}\nn\\
&-\frac{45G\mu_c}{4c^5}\frac{k^4a^3}{(1+kt)^3}-\frac{2796}{5}\frac{G^2M_c\mu_c}{c^5}k^2,
\end{align}
 with no closed-form solution. For slow mass loss, $kt\ll 1$, a post-Newtonian ordering (Appendix~\ref{AppCoalApprox}) shows that the leading early-inspiral balance is given by
\be\label{eqapp1}
\frac{\D a}{\D t}\apx-\bt\frac{(1+kt)^3}{a^3}+\frac{2ka}{1+kt},
\ee
where $\bt=64G^3M_{c1}M_{c2}M_c/5c^5$. The solution is
\begin{align}
a^4(t)=(1+kt)^4\lb a_0^4(1+kt)^4-\frac{\bt}{k}\lp(1+kt)^4-1\rp\rb.
\end{align}
Solving $a(T_c)=0$ gives
\be
T_c(a_0,0)=\frac{1}{k}\left[\lp\frac{\bt}{\bt-4a_0^4k}\rp^{1/4}-1\right],
\ee
which reduces to $T_c(a_0,0)\to a_0^4/(4\bt)$ as $k\to 0$.

\subsubsection{The exponentially varying mass case}
\color{black}

 If the mass variation rate is approximately directly proportional to the mass of the binary system, $\dot{M}\propto M$, then $f(t)$ is given by an exponential function. Hence we will assume that
\be
f(t)\apx{\rm e}^{\om t},
\ee
where $\omega $ is a constant.
Then Eq.~(\ref{vara}) for $(\D a/\D t)$  takes the form
\begin{align}\label{eqexp}
\frac{\D a}{\D t}=&-\frac{64G^3M_cM_{1c}M_{2c}}{5c^5}\frac{{\rm e}^{3\om t}}{a^3}+2a\om\nn\\
&-\lp492\frac{G^2\mu_cM_c}{c^5}\om^2{\rm e}^{2\om t}
+\frac{2817G\mu_c}{20c^5}\om^4{\rm e}^{\om t}a^3\rd\nn \\
&\ld+\frac{4\mu_c}{15c^5M_c}\om^6a^6\rp,
\end{align}
{which likewise has no closed analytical solution. The same slow-variation ordering (Appendix~\ref{AppCoalApprox}) reduces it to
\be
\frac{\D a}{\D t}\apx-\bt\frac{{\rm e}^{3\om t}}{a^3}+2a\om,
\ee
with solution
\begin{align}\label{eqaexp}
a^4(t)=\frac{4\bt{\rm e}^{3\om t}}{5\om}\lp1-{\rm e}^{5\om t}\rp+a_0^4{\rm e}^{8\om t}.
\end{align}
Solving $a^4(T_c)=0$ yields
\be
T_c(a_0,0)=\frac{1}{5\om}\ln\frac{4\bt}{4\bt-5\om a_0^4},
\ee
which again reduces to $a_0^4/(4\bt)$ as $\om\to 0$.

\subsubsection{The general linear time dependent mass function case}

From the analysis of these two special cases, we can find a more general conclusion: if the mass variation function can be expanded in the form that $f=1+kt+O(k)$, Eq.~(\ref{dadt}) for  $(\D a/\D t)$ and Eq.~(\ref{dedt}) for $(\D e/\D t)$ are approximately given by
\begin{align}
\left\langle\frac{\D a}{\D t}\right\rangle\apx&\frac{F_4^2}{f_1f_2}\lag\frac{\D a}{\D t}\rag_c+a\lp\frac{\dot{f_1}}{f_1}+\frac{\dot{f_2}}{f_2}\rp,
\end{align}
and
\begin{align}
\left\langle\frac{\D e}{\D t}\right\rangle\apx\frac{F_4^2}{f_1f_2}\left\langle\frac{\D e}{\D t}\right\rangle_c-\frac{(1-e^2)}{2e}\lp\frac{\dot{f}_M}{f_M}-3\frac{\dot{f}_1}{f_1}-3\frac{\dot{f}_2}{f_2}\rp,
\end{align}
respectively. When $f_1=f_2=f$, substituting $\left\langle\frac{\D e}{\D t}\right\rangle_c$ and $\left\langle\frac{\D e}{\D t}\right\rangle_c$, and omitting the symbol $\lag\rag$, we obtain
\begin{align}
\frac{\D a}{\D t}\apx&-\frac{\bt}{a^3\lp1-e^2\rp^{7/2}}\lp1+\frac{73}{24}e^2+\frac{37}{96}e^4\rp f^3+2a\frac{\dot{f}}{f},
\end{align}
\be
\frac{\D e}{\D t}=-\frac{19\bt e}{12a^4\lp1-e^2\rp^{5/2}}\lp1+\frac{121}{304}e^2\rp f^3+\frac{5(1-{\rm e}^2)}{2e}\frac{\dot{f}}{f}.
\ee

From the above equations we cannot obtain $a$ as a simple function of $e$, either. Considering the circular orbit, we have
\be\label{139}
\frac{\D a^4}{\D t}=-4\bt f^3+8a^4\frac{\dot{f}}{f}.
\ee

The general solution of Eq.~(\ref{139}) is given by
\be
a^4(t)=f^8\left(C-4\beta\int{\frac{dt}{f^5}}\right).
\ee

We can obtain $T_c(a_0,0)$ after solving the equation $a(0)=0$.

\section{Orbital decay due to the mass loss in binary magnetar systems}\label{sect4}

In the following we present a binary system simulator in which the influence of time varying masses is observed along with its effects on gravitational radiation. To do this, we will consider a system composed of two magnetars, which are neutron stars with a very strong magnetic field, representing an extreme class of compact objects relevant to the binary evolution discussed above. We will briefly explain the computational method used in such simulations, validate the program through the previous analytical cases of variable masses, and present a realistic astrophysical scenario in which the mass loss function is not analytically known. 

\subsection{The validity of quadrupole approximation}

The quadrupole formalism for gravitational radiation is a highly accurate approximation of general relativity. However, its validity is restricted to specific physical/astrophysical situations, requiring a slow motion of the source,  with internal velocities much smaller than the speed of light, $v \ll c$, and the weak gravitational field approximation,  with a small  gravitational potential energy $GM/c^2 R \ll 1$. Moreover, the quadrupole approximation is valid only in the radiation zone, where the distance to the observer is much larger than the size of the source \cite{Blanchet1}. The lowest-order quadrupole formula assumes that gravitational waves propagate through a flat, vacuum geometry, but non-linear scattering effects, like the scattering of the  waves by the  background curvature of the source, require additional corrections to the quadrupole formula.  In systems like coalescing black holes or neutron stars, the basic quadrupole formula breaks down during the late inspiral and merger phases \cite{Blanchet1}.  

To go  beyond the quadrupole approximation Post-Newtonian (PN) methods can be used, which consist in expanding the Einstein gravitational field equations in powers of $v/c$ \cite{Blanchet2}. The standard quadrupole formula represents the leading 2.5PN  term, and in order to explain the observational data gravitational wave astronomy must use calculations evaluated to 3.5PN. Generally, during the initial stages of the motion, the stars in a binary system  move in a weak gravitational field, where the PN approximation is valid. However, once the PN approximation criteria are not any longer valid, high-accuracy numerical templates based on full  general relativity theory are required \cite{Blanchet1,Blanchet2}.

After a certain amount of time, inspiralling compact binaries composed of neutron stars and/or black holes necessarily enter the strong field region. In this regime the methods based on perturbation theory cannot be applied anymore, implying that the description of the gravitational processes can only be done  within full numerical relativity \cite{Bini}. On the other hand in a coalescing process one can assume  that the particle may spiral for many cycles around the central compact object before entering in the strong field region.

Hence, in order to perform  numerical or analytical studies of gravitational radiation emission or of coalescence  based on Post-Newtonian approximations one must stop the PN description once the system approaches the strong gravitational field regime. This can be done by introducing a temporal scale  beyond which the PN formalism cannot be used anymore \cite{Bini}.
\color{black}

\subsection{Binary system simulations with GRAV-T}

We visualize the effects of the time variable mass system by evaluating the energy and angular momentum loss rates, given in Eqs.~\eqref{dE_mass} and  \eqref{dL_mass}, which influence the semimajor axis and eccentricity evolution in the most general form, Eqs.~\eqref{dadt} and \eqref{dedt}, respectively. We construct a binary system simulation through the GRAV-T program, which evaluates the time evolution of two point masses varying with respect to the function $f(t)$. In order to validate our program, we will assume for simplicity that the evolution function is analytical and follows a linear or exponential behaviour. The implementation evaluates the mass scaling function $f(t)$ and its derivatives up to third order, as required by the orbital evolution equations.

\paragraph{The numerical code.} The implemented code consists of a simple workflow.} Each simulation  is defined by a configuration file, written in a human readable format known as \verb|TOML|. This file contains the initial state configuration, namely the system's masses, semimajor axis and eccentricity, together with the decay loss rate of the function $f(t)$ and plotting parameters. The second stage evaluates the orbital parameters $a(t), e(t)$ and loss rates $-dE/dt, -dL/dt$ subject to the mass decay.

It must be noted that the evolution of the compact binary system presents numerical difficulties due to varying timescales between the early inspiral and the merger phase.  As the system evolves over a large time period, typically $10^{14}$ seconds, the large coalescence time $T_c(a, e)$ introduces a disparity between the time step size within the integration of the two stages. The orbital parameters change on timescales of years, yet near merger the characteristic timescale collapses to milliseconds. 

\paragraph{The computation of the coalescence time.} This issue is easily managed by introducing two separate integration runs on different time domains. The coalescence time $T_c$ is first computed using an event-detection algorithm that monitors when the semimajor axis reaches the innermost stable circular orbit (ISCO) $r_{\text{ISCO}} = 6GM/c^2$, with $M$ being the total mass of the system. The inspiral stage integrates from $t=0$ to $T_c - \Delta t_{\text{merger}}$ with only 10\% of the total integration points, having a time step of 10--100 years. The merger stage inherits the final state and integrates over the last few milliseconds with the remaining 90\% of points, achieving a temporal resolution of approximately $1~\mu$s per point. 
%
%\begin{figure}[htbp!]
%	\begin{center}
%		\scalebox{0.9}{
%		\begin{tikzpicture}[
%			node distance=1cm and 1.4cm,
%			classbox/.style={rectangle, draw, thick, minimum width=2.4cm, minimum height=0.8cm, font=\scriptsize, align=center, rounded corners=2pt},
%			configbox/.style={classbox, fill=blue!15, draw=blue!50},
%			corebox/.style={classbox, fill=green!15, draw=green!50},
%			plotbox/.style={classbox, fill=orange!15, draw=orange!50},
%			databox/.style={classbox, fill=yellow!20, draw=yellow!60},
%			arrow/.style={-{Stealth[scale=0.6]}, thick, draw=gray!70}
%			]
%			\node[configbox] (config) {\textbf{Config}\\TOML $\to$ params};
%			\node[configbox, right=of config] (state) {\textbf{State}\\$a$, $e$, $M_1$, $M_2$};
%			\node[corebox, below=0.5cm of config] (model) {\textbf{BinarySystem}\\$da/dt$, $de/dt$, \\ $dL/dt$, $dE/dt$};
%			\node[corebox, right=of model] (intrun) {\textbf{Integration}\\SciPy solver};
%			\node[databox, below=0.5cm of intrun] (h5) {\textbf{HDF5 Storage}};
%			\node[plotbox, below=0.5cm of h5] (odep) {ODEPlotter \\ MultiPlotter};
%			\draw[arrow] (config) -- (state);
%			\draw[arrow] (config) -- (model);
%			\draw[arrow] (state) -- (intrun);
%			\draw[arrow] (model) -- (intrun);
%			\draw[arrow] (intrun) -- (h5);
%			\draw[arrow] (h5) -- (odep);
%		\end{tikzpicture}
%	}
%	\end{center}
%	\caption{\textcolor{blue}{GRAV-T architecture showing data flows from configuration through physics and integration to HDF5 storage, from which plotters draw input.}}
%	\label{architecture}
%\end{figure}

Due to the large amount of integration steps in the simulation, we use the \verb|HDF5| file format, which provides gzip compression for efficient data storage. This format mimics a file system, allowing for hierarchical organization of data. We create a main HDF file, serving as a database for all the integration runs of a problem setup. Each integration run is stored as a separate group containing datasets for $a(t)$, $e(t)$, $M_1(t)$, $M_2(t)$, and the corresponding time array.

\paragraph{Output of the results.}  The last stage manages and creates plots using the data stored in the HDF files. The orbital parameters' evolution throughout the inspiral phase is displayed together with comparative plots overlaying results from multiple decay rates, as presented in the figures presented in what follows.\color{black}

\subsection{Validation with existing PN software}
	
The presented numerical framework was subject to comparison with a pre-existing PN simulation software, LEGWORK \cite{legwork}, which provides an easy to handle interface for simulating the evolution of binary systems. However, as GRAV-T is currently the only implementation capable of simulating gravitational radiation effects from variable mass systems, the comparison between our software and LEGWORK is done in the constant mass case, where we recover the Peter-Mathews results. For this validation, both codes were evolved from identical initial conditions, with mass-loss switched off in GRAV-T so that its equations reduce to the standard Peters-Mathews limit. GRAV-T was integrated as a coupled system for the semi-major axis and eccentricity up to its coalescence time, defined by the orbit reaching the ISCO. In parallel, LEGWORK was used through its eccentric inspiral module, which implements the same Peters-Mathews equations on a simulation grid up to a precalculated merger time. 

The two trajectories were compared by interpolating the GRAV-T solution onto the same uniform time mesh used for LEGWORK, evaluating the relative differences
$\Delta a_{\text{rel}} = (a_{\text{GT}} - a_{\text{LW}})/a_{\text{LW}}$ and
$\Delta e_{\text{rel}} = (e_{\text{GT}} - e_{\text{LW}})/e_{\text{LW}}$.
Because the absolute residual between the two codes remains at the sub-percent level throughout the inspiral, a direct overlay of $a_{\text{GT}}(t)$ with $a_{\text{LW}}(t)$ (and likewise for $e$) would be visually indistinguishable, therefore the quantitative validation is presented through the relative residuals themselves.

\begin{figure}[htbp!]
	\centering
	\includegraphics[scale=0.7]{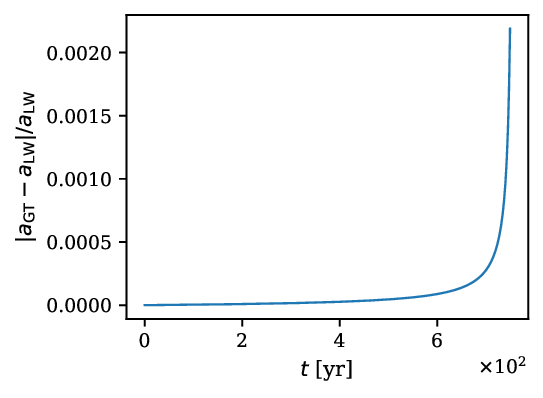}
	\includegraphics[scale=0.7]{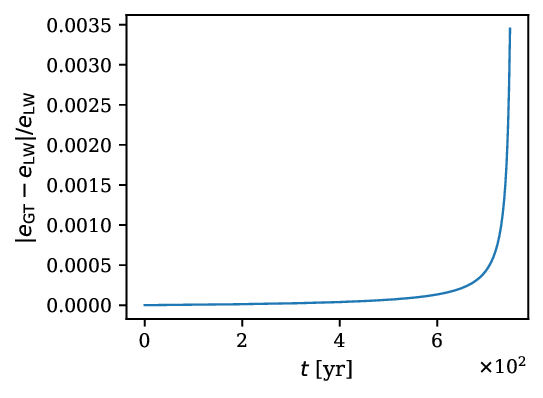}
	\caption{The relative errors in semi-major axis (left panel) and eccentricity evolution (right panel) throughout GRAV-T and LEGWORK codes.}
	\label{benchmark}
\end{figure}

We report the minimum, mean, and maximum of these residuals over the common time window in Table~\ref{errel}, and their full time dependence in Fig.~\ref{benchmark}.
Both panels of Fig.~\ref{benchmark} share the same horizontal time axis, with the vertical scales differing only because $\Delta a_{\text{rel}}$ and $\Delta e_{\text{rel}}$ are distinct dimensionless residuals and need not have the same amplitude.
The mean relative discrepancies are $\langle\Delta a_{\text{rel}}\rangle\sim 8\times10^{-5}$ and $\langle\Delta e_{\text{rel}}\rangle\sim 1\times10^{-4}$, with maxima of order $10^{-3}$ only near merger, where the post-Newtonian description becomes increasingly stiff.
This confirms that GRAV-T reproduces the Peters-Mathews dynamics before mass-variation effects are considered.
\color{black}

\begin{table}[htbp!]
	\centering
	\begin{tabular}{|l|c|c|c|}
		\hline
		 & Min & Mean & Max \\
		\hline
		$\Delta a_{\text{rel}}$ & 5.692e-16 & 8.305e-05 & 2.189e-03 \\
		$\Delta e_{\text{rel}}$ & 0.000e+00 & 1.279e-04 & 3.450e-03 \\
		\hline
	\end{tabular}
	\caption{Relative residuals between GRAV-T and LEGWORK evaluated on a common time mesh.}
	\label{errel}
\end{table}

\subsubsection{Linear mass decay}

Within the aforementioned simulation setup, we considered a typical neutron star binary system of equal masses $M_{1c}=M_{2c}=1.4M_\odot$, with time varying functions associated to their mass decay given by $f_1(t)=f_2(t)=f(t)$. As a first test we adopt the linear analytical function $f(t)=1-kt$, where $k$ is a positive coefficient describing the mass loss rate. To visualize the effects of rapid and slow mass variations, we use $k=\{10^{-15},10^{-13},2\times10^{-13},3\times10^{-13}\}$ s$^{-1}$, chosen so that the coalescence time remains of the order of thousands of years and the stars still merge before the mass loss becomes disruptive.

\begin{figure*}[htbp!]
	\centering
	\includegraphics[scale = 0.45]{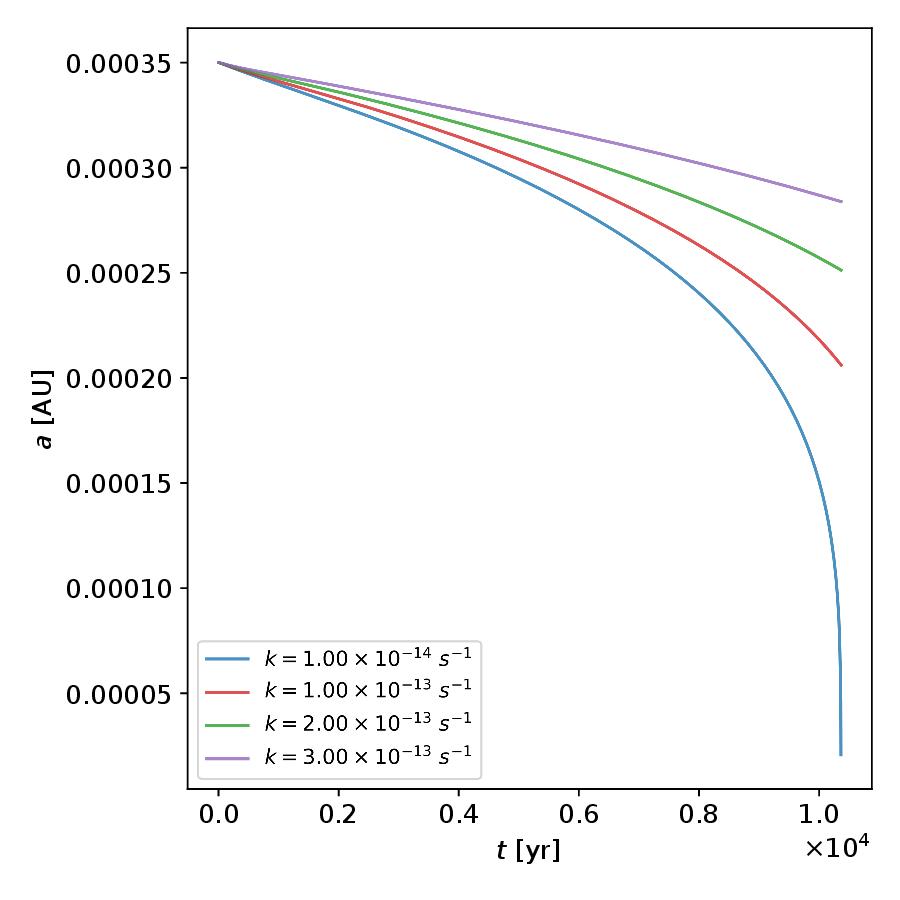}
	\includegraphics[scale = 0.45]{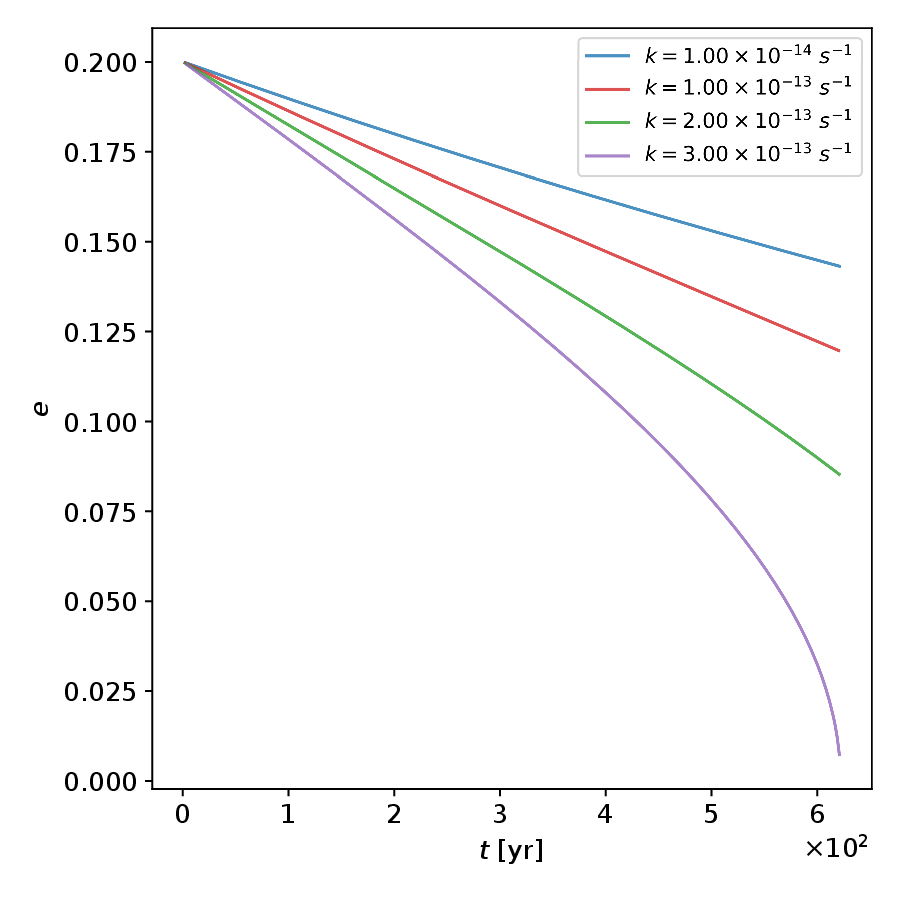}
	\caption{Orbital evolution for four linear mass decay rates during the inspiral phase. The semimajor axis decay (left panel) shows the characteristic inspiral trajectory toward ISCO and the eccentricity evolution (right panel) shows circularization.}
	\label{odeplots}
\end{figure*}

\begin{figure*}[htbp!]
	\centering
	\includegraphics[scale = 0.38]{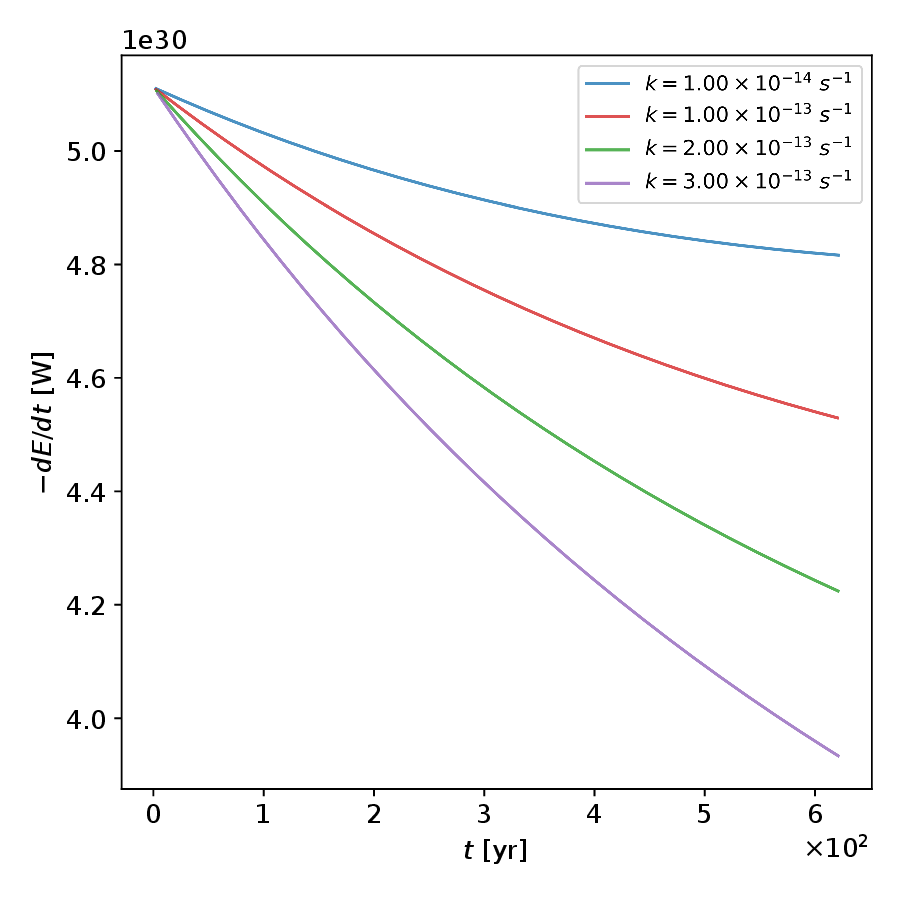}
	\includegraphics[scale = 0.38]{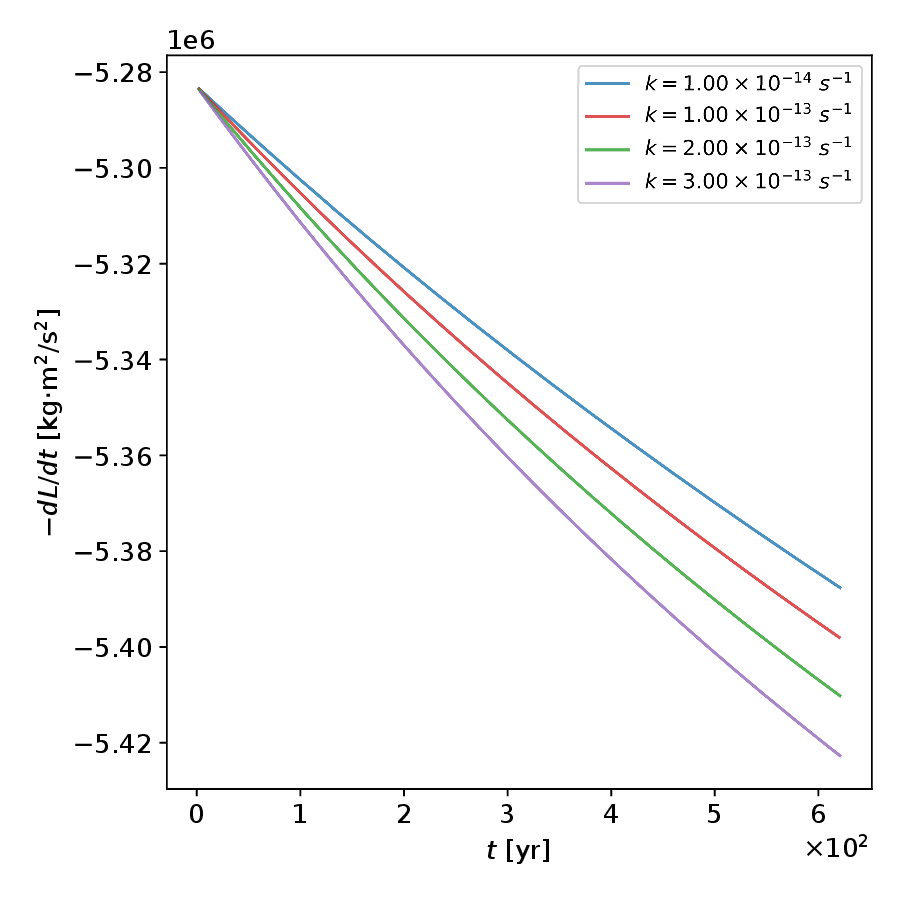}
	\includegraphics[scale = 0.38]{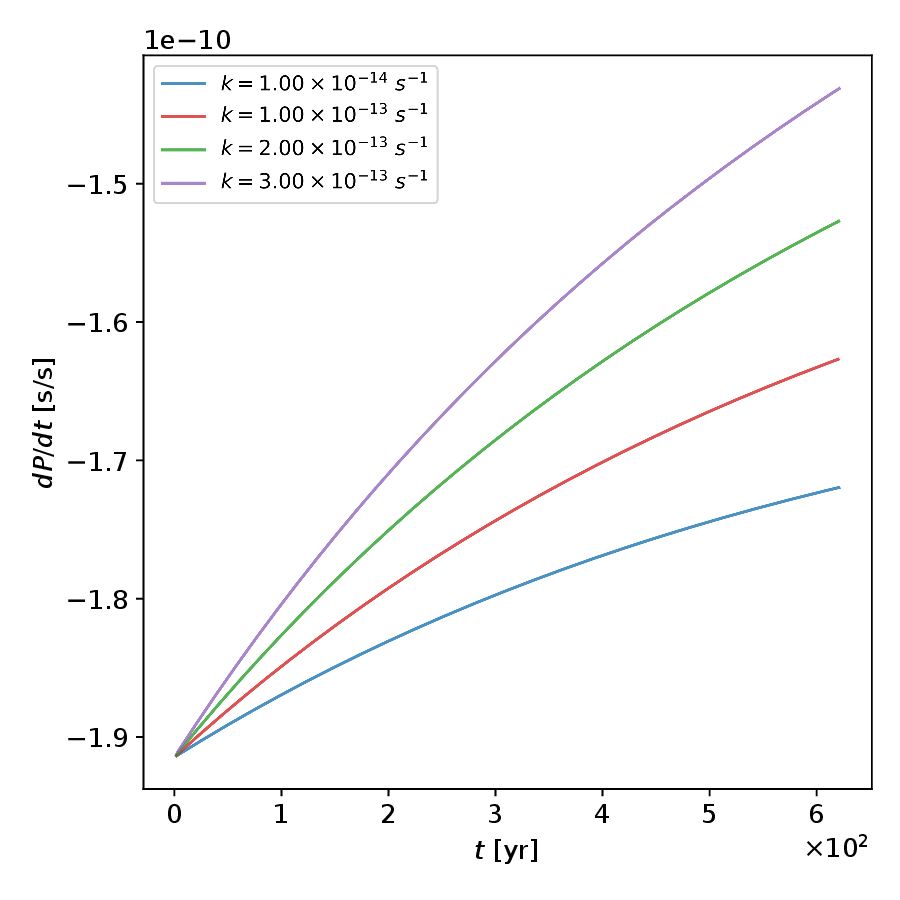}
	\caption{Comparison of energy (left panel), angular momentum (middle panel), and orbital period loss (right panel) rates for different linear mass decay parameters.}
	\label{ELPplots}
\end{figure*}

While typical neutron star binaries have lifetimes of $10^6-10^9$ years, we initialize the system at a late stage because linear mass loss can destabilize the binary before merger. By setting a separation of $a_0=3.5\times10^{-4}$ AU $\approx 5.2\times10^4$ km, the corresponding coalescence time is $T_c\approx 3\times10^{11}$. The system begins with eccentricity $e_0=0.2$, and the simulation is considered complete at the separation $r_{\text{ISCO}}$. The evolution of the orbital elements, together with the energy, angular momentum, and orbital-period variations, is displayed in Figs.~\ref{odeplots} and \ref{ELPplots}. For direct comparison, the curves are clipped to the time domain of the fastest coalescence, and the corresponding coalescence times are summarized in Table~\ref{tab:linear_tc}.

\begin{table}[htbp!] 
	\centering 
	\begin{tabular}{|c|c|c|} 
		\hline 
		$k$ [s$^{-1}$] & $T_c$ [s] & $T_c$ [years] \\ 
		\hline 
		$1.00 \times 10^{-15}$ & $3.23 \times 10^{11}$ & $1.02 \times 10^{4}$ \\ 
		$1.00 \times 10^{-13}$ & $3.72 \times 10^{11}$ & $1.18 \times 10^{4}$ \\ 
		$2.00 \times 10^{-13}$ & $4.45 \times 10^{11}$ & $1.41 \times 10^{4}$ \\ 
		$3.00 \times 10^{-13}$ & $5.92 \times 10^{11}$ & $1.88 \times 10^{4}$ \\ 
		\hline 
		\end{tabular} 
	\caption{Coalescence time for different linear decay rates.} 
	\label{tab:linear_tc} 
\end{table}

The semimajor axis $a(t)$ (Fig.~\ref{odeplots}, left panel) describes the characteristic inspiral trajectory, decreasing from its initial value of $3.5\times10^{-4}$ AU toward $r_{\text{ISCO}}\approx 28.4$ km. The lowest decay rate (purple, $k=10^{-15}$ s$^{-1}$) reaches $r_{\text{ISCO}}$ first, while higher decay rates result in longer coalescence times as the reduced mass slows down the inspiral. The eccentricity evolution (Fig.~\ref{odeplots}, right panel) shows a rapid circularization from $e_0=0.2$ for the highest decay rate (yellow, $k=3\times10^{-13}$ s$^{-1}$), so that $e$ reaches zero before the simulation ends. For this reason the time frame for the eccentricity, and consequently for the quantities shown in Fig.~\ref{ELPplots}, ends at $t=10^{10}$ s, the moment when the orbit circularizes.

The energy loss rate $-dE/dt$ (Fig.~\ref{ELPplots}, left panel) ranges from approximately $5.1\times10^{30}$ W to $3.8\times10^{30}$ W depending on the decay rate. The luminosity increases sharply as the orbit shrinks, following the term $F_4^2\langle dE/dt\rangle_c$ of Eq.~\eqref{dE_mass}, with $\langle dE/dt\rangle_c\propto a^{-5}$. The curves for higher mass loss rates are slightly lower in magnitude, as expected for a lighter binary system. The angular momentum loss rate $-dL/dt$ (Fig.~\ref{ELPplots}, middle panel) displays a nearly linear decrease for small decay rates and a more sublinear shape for high decay rates, ranging from approximately $-4.75\times10^{11}$ to $-4.84\times10^{11}$ kg m$^2$/s$^2$ as the system evolves. The orbital period derivative $dP/dt$ (Fig.~\ref{ELPplots}, right panel) varies from approximately $-1.18\times10^{-10}$ to $-0.7\times10^{-10}$ and remains negative, confirming the continued shrinkage of the orbit.

\subsubsection{Exponential mass decay}

The other simulation constructed for the program validation contains the exponentially varying mass function $f(t)=e^{-\omega t}$, with $\omega>0$, again considering a system of two neutron stars with equal masses $M_{1c}=M_{2c}=1.4M_\odot$. This second problem is initialized with a semimajor axis $a_0=10^{-3}$ AU $\approx 1.5\times10^5$ km and eccentricity $e_0=0.1$. The wider initial separation results in a significantly longer inspiral phase than in the linear case, with coalescence times of $2.2\times10^{13}$ s. The decay parameters are chosen in the range $\omega\in\{2\times10^{-17},2\times10^{-15},4.0\times10^{-15},6.0\times10^{-15}\}$ s$^{-1}$ to ensure numerical stability over this longer evolution.

\begin{figure*}[htbp!]
	\centering
	\includegraphics[scale = 0.45]{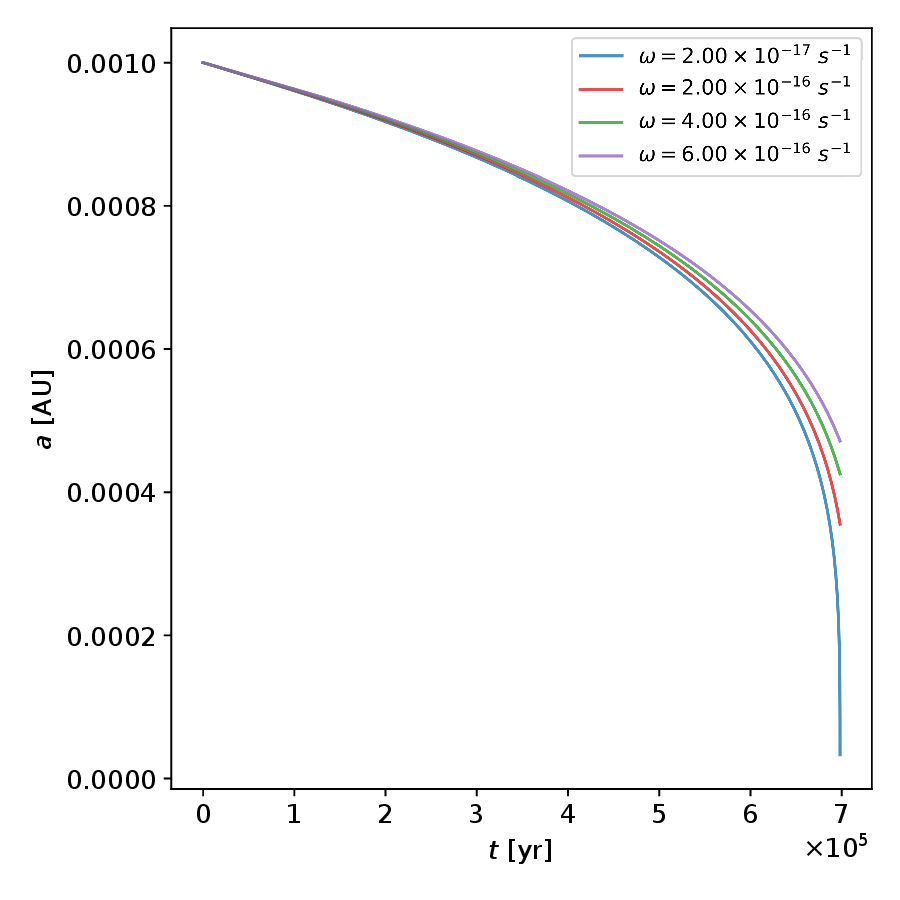}
	\includegraphics[scale = 0.45]{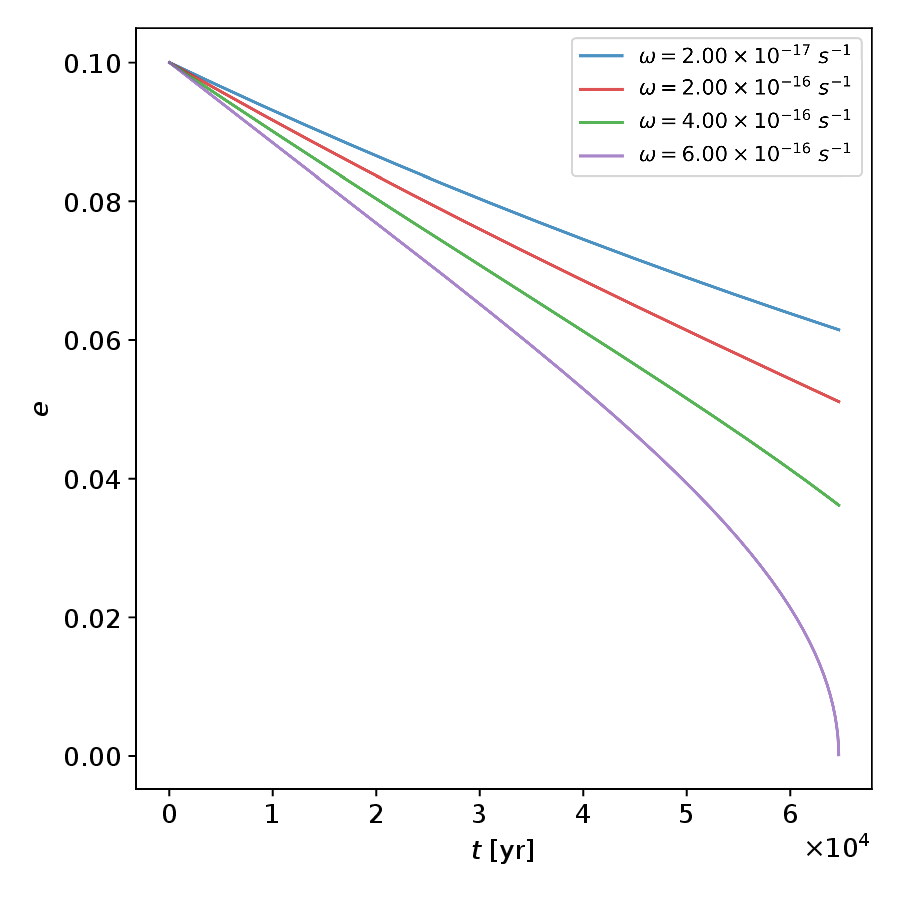}
	\caption{The semimajor axis (left panel) and eccentricity evolution (right panel) for the binary system with exponential time dependent mass decay.}
	\label{odeplots_exp}
\end{figure*}

\begin{figure*}[htbp!]
	\centering
	\includegraphics[scale = 0.38]{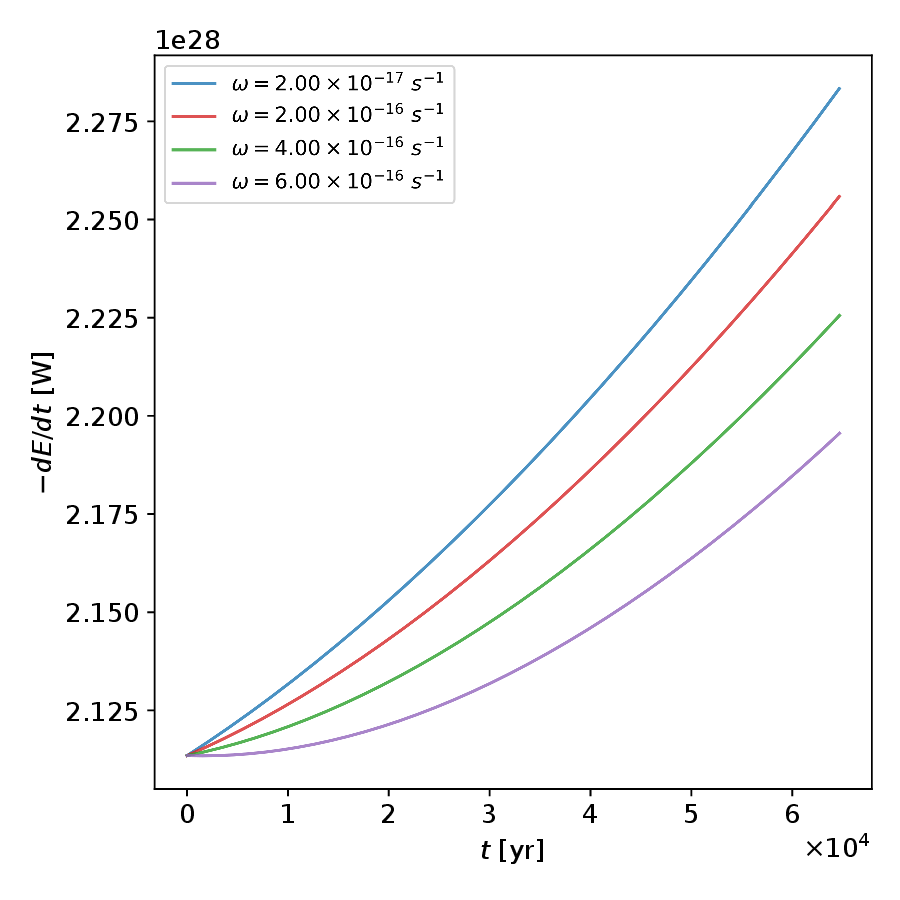}
	\includegraphics[scale = 0.38]{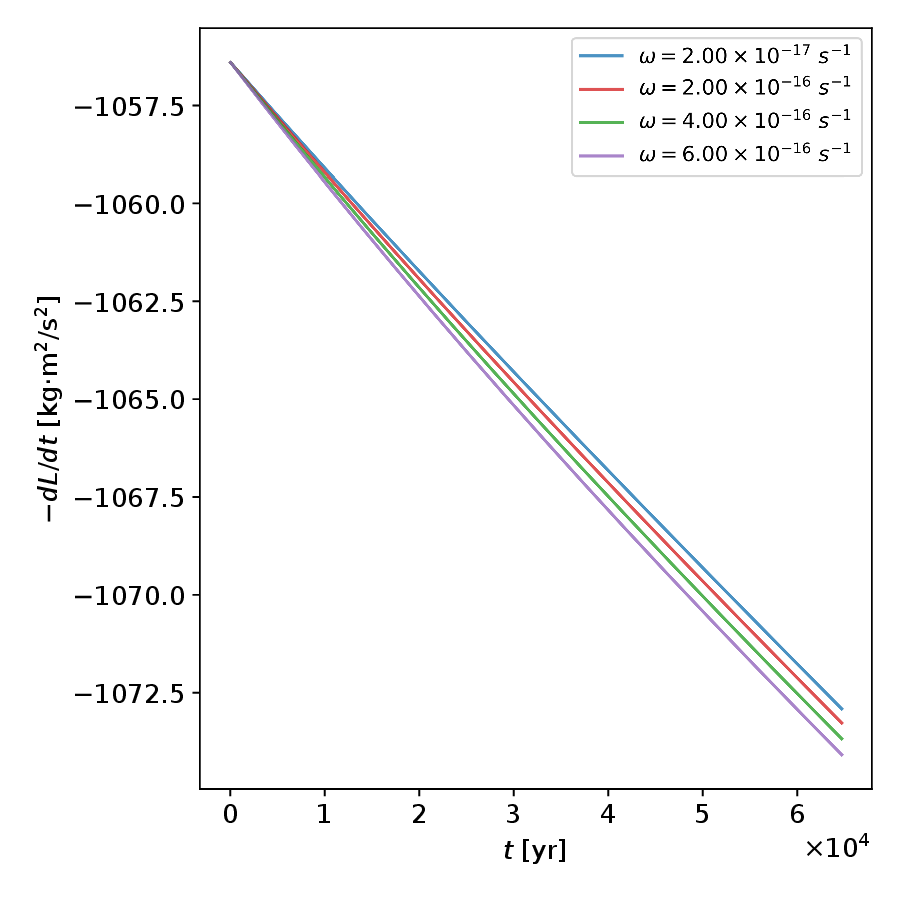}
	\includegraphics[scale = 0.38]{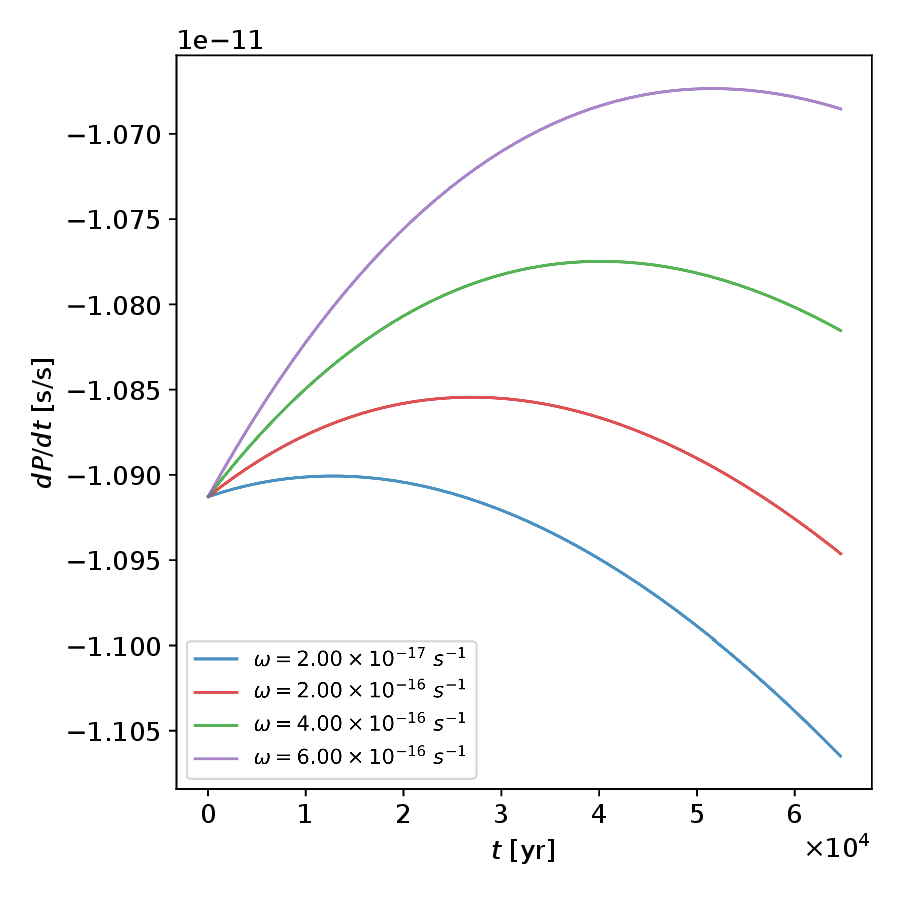}
	\caption{Energy decay (left panel), angular momentum loss (middle panel) and orbital period decay (right panel) for the binary system with exponential mass decay.}
	\label{ELPplots_exp}
\end{figure*}

The evolution of the orbital elements and conservation laws is presented in Figs.~\ref{odeplots_exp} and \ref{ELPplots_exp}. As in the linear case, all simulations reach ISCO, but the visualization is  clipped to the time domain of the fastest coalescence ($\omega=2\times10^{-17}$ s$^{-1}$). The corresponding coalescence times are given in Table~\ref{tab:exp_tc}.

\begin{table}[htbp!] 
	\centering 
	\begin{tabular}{|c|c|c|} 
		\hline 
		$\omega$ [s$^{-1}$] & $T_c$ [s] & $T_c$ [years] \\ 
		\hline 
		$2.00 \times 10^{-17}$ & $2.20 \times 10^{13}$ & $6.98 \times 10^{5}$ \\ 
		$2.00 \times 10^{-15}$ & $2.65 \times 10^{13}$ & $8.39 \times 10^{5}$ \\ 
		$4.00 \times 10^{-15}$ & $3.45 \times 10^{13}$ & $1.09 \times 10^{6}$ \\ 
		$6.00 \times 10^{-15}$ & $6.34 \times 10^{13}$ & $2.01 \times 10^{6}$ \\ 
		\hline 
	\end{tabular} 
	\caption{Coalescence time for exponential decay rates.} 
	\label{tab:exp_tc} 
\end{table}

The semimajor axis $a(t)$ (Fig.~\ref{odeplots_exp}, left panel) shows a smooth decay from $10^{-3}$ AU, with the lowest decay rate (purple) reaching $r_{\text{ISCO}}$ first, as in the linear case. Higher decay rates result in longer coalescence times. The eccentricity evolution (Fig.~\ref{odeplots_exp}, right panel) reveals the fastest circularization for the highest decay rate (yellow, $\omega=6\times10^{-15}$ s$^{-1}$) within $t=10^{12}$ s, while lower rates show only partial circularization at the clip boundary. All orbits nevertheless circularize within the coalescence time $T_c$ evaluated at $r_{\text{ISCO}}$.

The energy loss rate $-dE/dt$ (Fig.~\ref{ELPplots_exp}, left panel) increases in magnitude from approximately $2.11\times10^{28}$ W to $2.27\times10^{28}$ W over the integration window. Higher decay rates result in lower radiated power, as the effective mass decreases more rapidly. This contrasts with the linear case, where the energy loss rate decreases over time, and it reflects the competition between the standard GW inspiral, which increases the luminosity as $a^{-5}$, and the mass loss, which decreases it through the mass dependence of the flux. In the exponential simulations the decay parameters are much smaller than in the linear case, so the gravitational-wave dynamics dominate for most of the inspiral. The angular momentum loss rate $-dL/dt$ (Fig.~\ref{ELPplots_exp}, middle panel) decreases from approximately $-9.54\times10^{5}$ to $-9.66\times10^{5}$ kg m$^2$/s$^2$, with closely grouped curves. The orbital period derivative $dP/dt$ (Fig.~\ref{ELPplots_exp}, right panel) ranges from approximately $-1.00\times10^{-11}$ to $-1.035\times10^{-11}$. For the lowest decay rate the curve shows an almost monotonic decrease, while for higher decay rates it turns upward slightly at late times as mass loss starts to compete with gravitational radiation.

\subsection{Mass loss from neutrino-heated winds in magnetars}

Magnetars lose a significant amount of mass through strong neutrino-heated winds, which provide a concrete astrophysical application of the formalism developed above. According to Lander and Jones \cite{Lander}, the mass-loss rate of a non-rotating, demagnetised proto-neutron star with initial mass $M=1.4M_\odot$ can be written as
\begin{gather}
	\dot{M}=-6.8\times10^{-5}M_\odot{\rm s}^{-1}\lp\frac{L_\nu}{10^{52}{\rm erg\ s^{-1}}}\rp^{5/3}\lp\frac{E_\nu}{10{\rm MeV}}\rp^{10/3}, \nn \\ \label{Lander1}
\end{gather}
where $M_\odot=1.9891\times10^{30}\; \rm kg$ is the solar mass, and the neutrino luminosity and energy per neutrino are expressed as
\begin{gather}
	\frac{L_\nu}{10^{52}{\rm erg\ s^{-1}}}\apx0.7\exp\lp-\frac{t[{\rm s}]}{1.5}\rp+0.3\lp1-\frac{t[{\rm s}]}{50}\rp^4,\nn\\
	\frac{E_\nu}{10{\rm MeV}}\apx0.3\exp\lp-\frac{t[{\rm s}]}{4}\rp+1-\frac{t[{\rm s}]}{60}. \label{Lander2}
\end{gather}

We now apply this mass-loss rate to analyse the orbital evolution described in Section~\ref{variable_mass}. For a binary system consisting of two identical magnetars, we assume initial masses $M_{1c}=M_{2c}=1.4M_\odot$, giving a total system mass $M_c=2.8M_\odot$ and reduced mass $\mu_c=0.7M_\odot$, so that
\be
\dot{f}=\frac{\dot{M}}{2\cdot 1.4M_\odot}.
\ee
Unlike the analytical test cases presented above, the mass scaling function $f(t)=1+\int_0^t\dot{f}(t')dt'$ must be evaluated numerically using quadrature at each integration step. The implementation computes the derivatives needed in the modified GW equations by differentiating the neutrino luminosity and energy expressions. The system is initialized with separation $a_0=5\times10^{-4}$ AU and eccentricity $e_0=0.2$. To investigate the dependence on the initial mass configuration, we vary the primary mass over $M_1\in\{1.2,1.4,1.6\}M_\odot$ while keeping the secondary mass fixed at $M_2=1.4M_\odot$.

\begin{figure*}[htbp!]
	\centering
	\includegraphics[scale = 0.45]{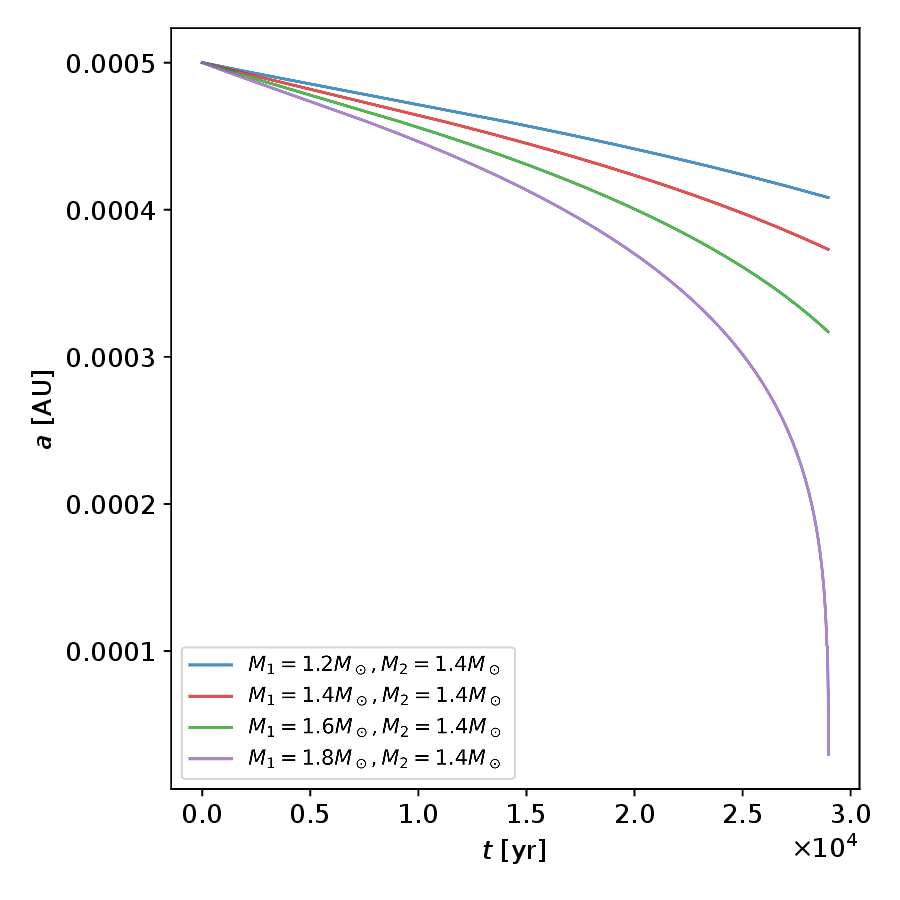}
	\includegraphics[scale = 0.45]{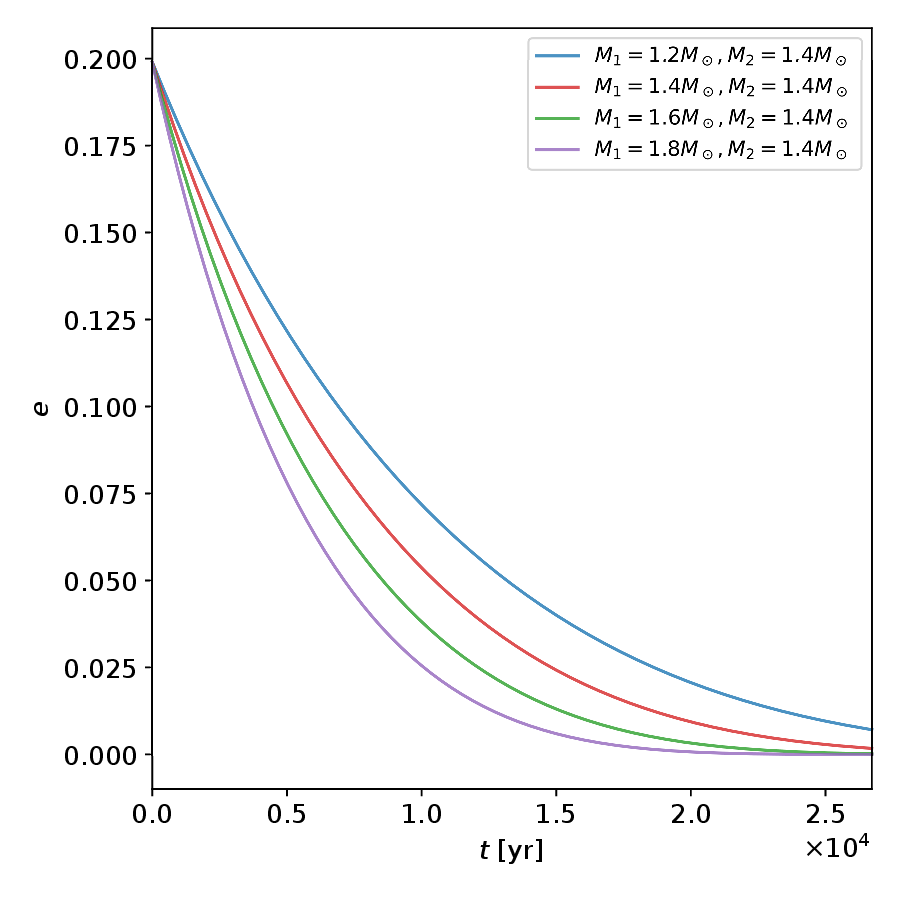}
	\caption{Orbital evolution for the Lander-Jones neutrino-heated wind model with varying primary mass $M_1$. The semimajor axis evolution (left panel) shows mass-dependent coalescence times. and the eccentricity circularization (right panel) shows strong dependence on the total system mass.}
	\label{odeplots_lander}
\end{figure*}

\begin{figure*}[htbp!]
	\centering
	\includegraphics[scale = 0.38]{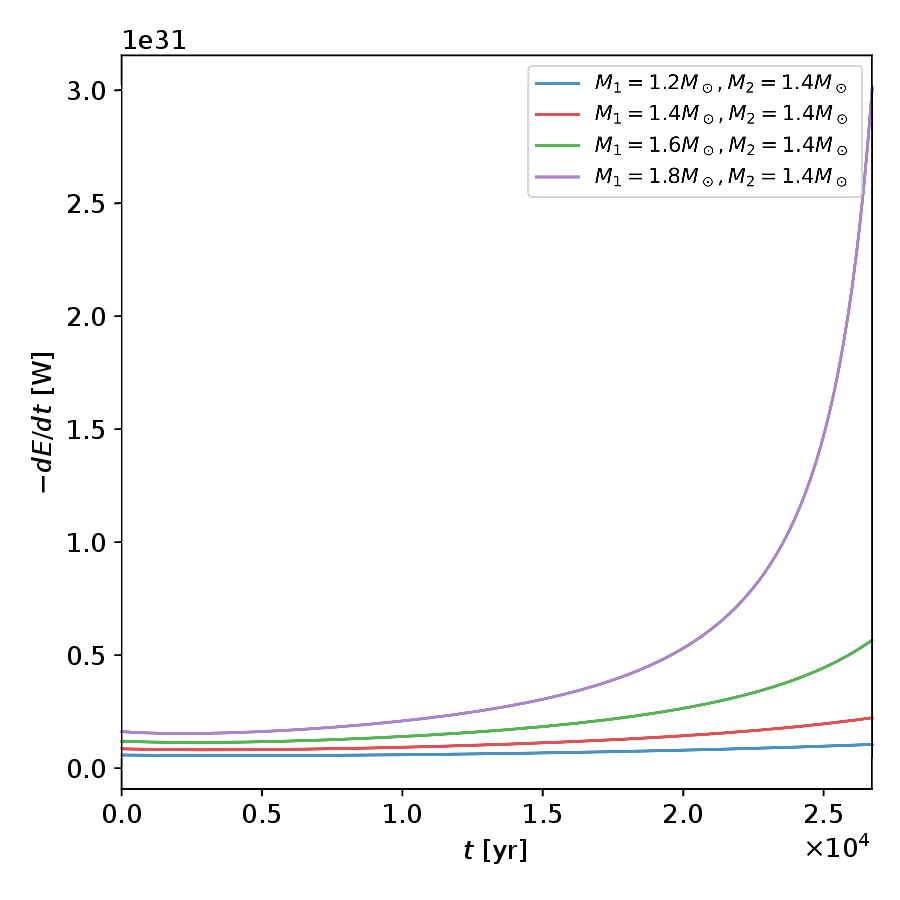}
	\includegraphics[scale = 0.38]{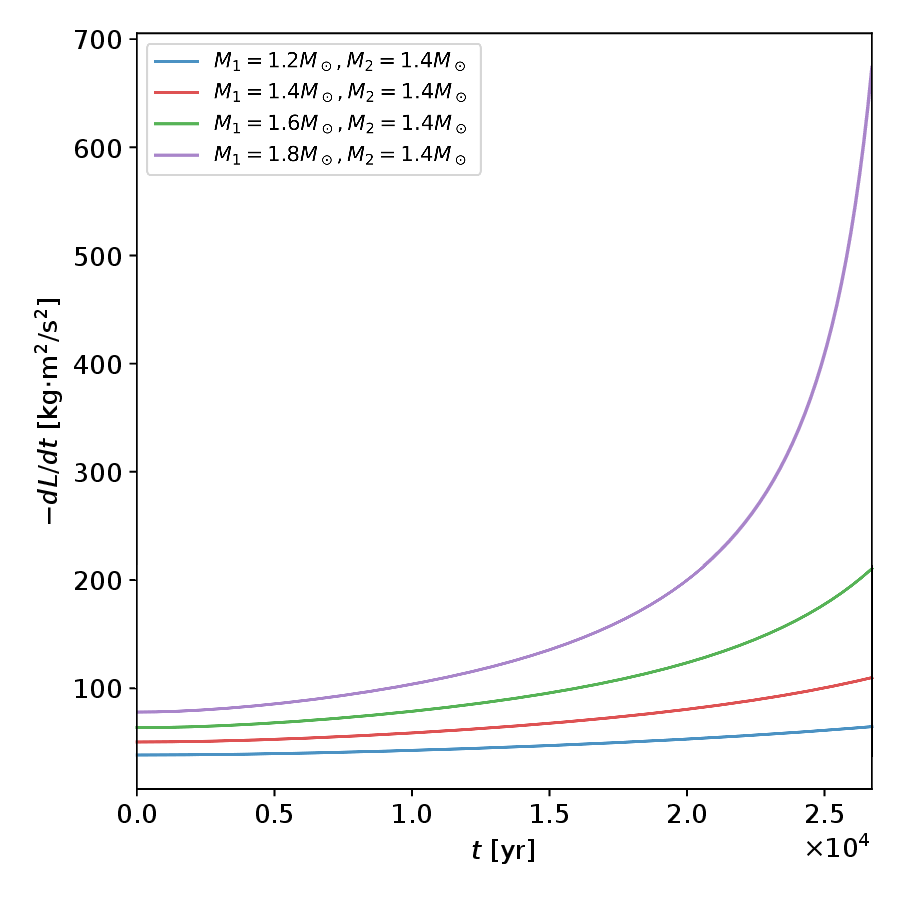}
	\includegraphics[scale = 0.38]{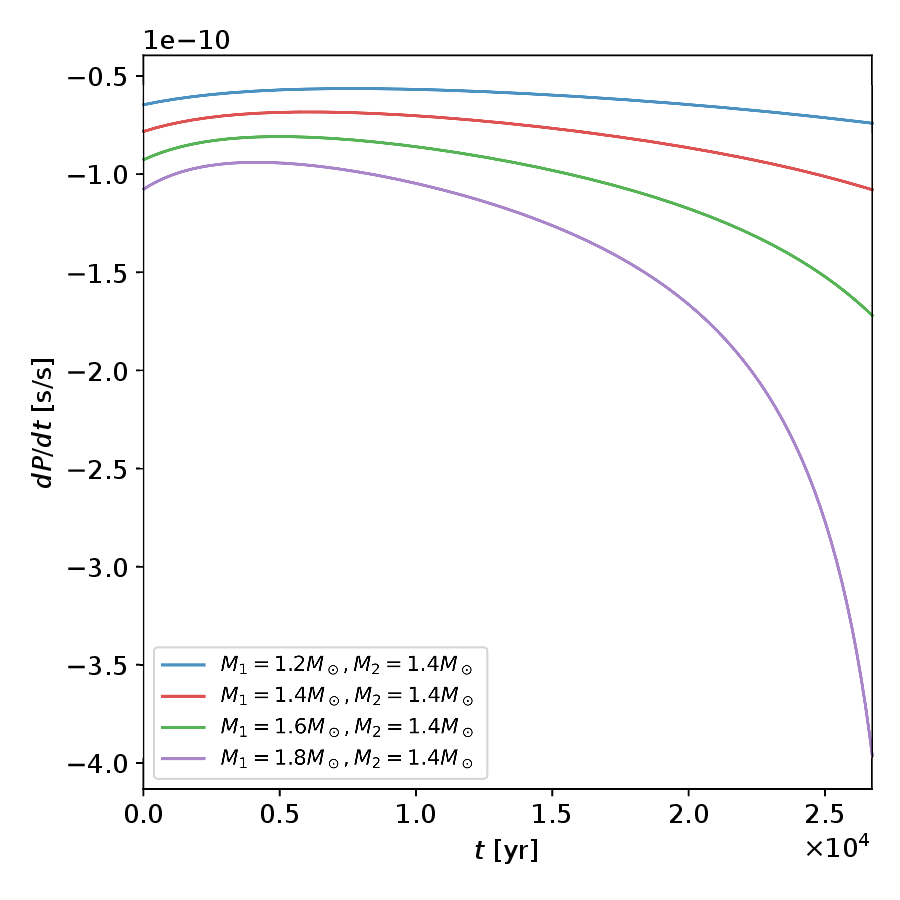}
	\caption{Energy decay (left panel), angular momentum loss rates (middle panel) and orbital period decay rate (right panel) for the Lander-Jones model with varying $M_1$. The curves show the strong mass dependence of gravitational radiation efficiency, with the heavier systems radiating more power but also losing angular momentum more rapidly.}
	\label{ELPplots_lander}
\end{figure*}

The semimajor axis evolution (Fig.~\ref{odeplots_lander}, left panel) reveals that higher masses in the system result in faster inspiral and shorter coalescence times, as expected from the stronger gravitational-wave emission at larger mass. The coalescence time is $T_c\sim10^{12}$ s, presented in Table~\ref{tab:lander_tc}. The eccentricity evolution (Fig.~\ref{odeplots_lander}, right panel) shows rapid circularization in all cases, with lighter systems circularizing somewhat faster because of the fractional mass-loss rate, $\dot{f}/f=\dot{M}/(2Mf)$.

\begin{table}[htbp!] 
	\centering 
	\begin{tabular}{|c|c|c|} 
		\hline 
		$M_1/M_2$ & $T_c$ [s] & $T_c$ [years] \\ 
		\hline 
		$0.086$ & $1.68 \times 10^{12}$ & $5.32 \times 10^{4}$ \\ 
		$1.000$ & $1.34 \times 10^{12}$ & $4.25 \times 10^{4}$ \\ 
		$1.142$ & $1.09 \times 10^{12}$ & $3.46 \times 10^{4}$ \\ 
		\hline 
	\end{tabular} 
	\caption{Coalescence time for different mass ratios in the neutrino-heated winds model.} 
	\label{tab:lander_tc} 
\end{table}

The energy loss rate (Fig.~\ref{ELPplots_lander}, left panel) ranges from approximately $0.5\times10^{30}$ W for the lightest system to $2.2\times10^{31}$ W for the heaviest, showing that the gravitational-wave luminosity scales strongly with the system mass. The angular momentum loss rate (Fig.~\ref{ELPplots_lander}, middle panel) follows a similar trend, reaching about $600$ kg m$^2$/s$^2$ for the heaviest case, while the orbital period derivative (Fig.~\ref{ELPplots_lander}, right panel) reaches $-4\times10^{-10}$ at $t=10^{12}$ s. Compared with the exponential decay case, the separation between curves is much stronger because varying $M_1$ changes both the total and reduced masses of the binary. A further difference is that the neutrino-heated wind mass loss is concentrated in the first $\sim 60$ s, after which the system evolves essentially as a constant-mass binary under standard gravitational-wave emission, unlike the linear and exponential models where the correction terms remain active throughout the integration.

\section{Discussions and final remarks}\label{sect5}

In the present paper, we have considered the theory of the generation of the gravitational waves  from binary system by considering its extension for time varying mass systems. We have reformulated  the general theory of gravitational radiation emission from binary systems with time variable gravitational mass in a different and more compact matrix form, as compared to the previous investigations \cite{Cheng1, Holgado1, Holgado2}. In our general approach we have not assumed any particular form for the mass variation. By using the quadrupole formalism we have obtained the full expressions of the time variation of the energy and momentum losses through gravitational radiation, as well as the dynamical evolution of the semimajor axis, eccentricity and period of the binary system with varying mass. Our results can lead to a better understanding of the theoretical foundations of gravitational radiation emission in binary systems, and to the observational phenomena related to these processes. In the present study we have provided the expressions of the relevant observational quantities that can be used to test in realistic astrophysical environments the validity of the quadrupole formalism of general relativity.

In our approach we have investigated  the superposition  of two purely general relativistic effects, the gravitational radiation, described through the quadrupole formalism, and the time dependence of the gravitational mass of compact objects, which  can be the result of several physical processes. The decay of the spin of the pulsars due to, for example, the magnetic dipole radiation) determines a change in the gravitational mass of the rotating object. When this effect is integrated in the quadrupole formalism of the gravitational radiation,
some supplementary terms in the classical formula of Peters and Mathews \cite{Peters, Peters1, Hansen} do appear. The Peters and Mathews formalism  has been tested successfully for the case of PSR 1913+16 \cite{Taylor, Weis} and other binary pulsars \cite{Fonseca}.

As a direct  astrophysical applications of our results  we have investigated the  coalescence time for time varying mass compact objects in two special cases,  by adopting two simple models for the gravitational mass decay. First, we have assumed that for a circular orbit, the mass of both the binary stars vary by a linear function,  $M(t)\propto 1+kt+O(k)$. In a second model we have assumed that the variation of mass is given by  an exponential function, respectively, with $M_i(t)\propto M_{i0}\exp (-\omega t)$, $i=1,2$. From our analysis, it follows that in the early period of evolution of the binary system the orbital decay may be influenced mainly by the mass variation rather than gravitational radiation. As an example of a specific astrophysical system  we have considered  the case  when both of the binary stars are magnetars losing mass due to the strong neutrino-heated wind of charged particles.   

In our present investigation we have assumed that, as a result of certain astrophysical processes like, for example, the presence of stellar winds or spinning down due to emission of electromagnetic radiation, the masses of the bodies vary adiabatically, implying a very slow mass variation. In a more precise formulation we have assumed that the time change of the masses of the two stellar objects during one period $T$ of the motion is very small, and it satisfies the important condition $T dM_i/dt<< M_i $, $i=1,2$. 
It is a well known result in theoretical physics that when the parameters of the motion change slowly, there are several constants of motion, called adiabatic invariants. If the physical system can be described by a Hamiltonian $H = H(q, p, \lambda$), where by $q$ we have denoted the generalized coordinate, $p$ is the momentum associated to $q$, and $\lambda$ is the slowly varying
parameter, the adiabatic invariant is given by the quantity $J=\frac{1}{2\pi}\oint{pdq}$. Hence, when the time variation of $\lambda $ is small, $J$ remains constant \cite{Landau1}.

For the case of the motion of two objects with varying mass, interacting via the gravitational force, and moving in a Keplerian orbit, there are two adiabatic invariants. The first is $J_\theta = (1/2\pi)\oint{p_{\theta}d\theta }=L$,  where $L$ denotes the total angular momentum of the system. Therefore, even though the masses of the astrophysical objects vary in time, $L$ remains a constant of motion.

As for the second adiabatic invariant of the motion it can be obtained as \cite{Landau1}
\be
J_r=\frac{1}{2\pi}\int_{r_{min}}^{r_{max}}{p_rdr}=-L+GM_1(t)M_2(t)\sqrt{\frac{\mu (t)}{2|E|}},
\ee
 where $\mu(t)$ is the reduced mass of the system. Therefore, as a function  of the
action variables, the total energy of the two body system is given  
\be
E = -\frac{G^2M_1^2(t)M_2^2(t) \mu(t)}{2\left(J_r+
J_\theta \right)^2}.
\ee 
Moreover, the eccentricity $e$ of the orbit can be obtained according to
\be
e^2 = 1-\left( \frac{J_\theta}{J_r+J_\theta}\right)^2. \label{e}
\ee
Hence, by taking into account that $J_r$ and $J_\theta$ are adiabatic invariants, from the above relation it follows that if the mass variation of the system is slow, then the eccentricity of the orbit does not change in time.

The variation of the semimajor axis of the orbit $a$ is given by 
\be
a =\frac{ \left(J_r+J_\theta \right)^2}{GM_1(t)M_2(t)\mu (t) }.
\ee
Hence it follows that as  a result of the time variation of the masses of the binary system, the semimajor axis of the orbit is also changing in time.

In the action-angle formalism the period of the motion is given, by the third Kepler law as
\be
P_b =2\pi \left(\frac{ \partial E}{ \partial J_r}\right)^{-1}= 2\pi \frac{ a^{3/2}}{\sqrt{G\left(M_1(t)+M_2(t)\right)}}. 
\ee
Therefore from the above relations it follows that the orbits of the  Newtonian motion of two bodies with slowly varying masses  are closed curves, and the motion is periodic. Hence, in the Newtonian approximation, for a binary system with time varying masses, the only source of time variation of the angular momentum $L$ and eccentricity $e$ is represented by the gravitational radiation. On the other time, the period of the orbit, the semi-major axis of the trajectory, and the energy is also influenced by gravitational radiation emission. 

The conservation of $e$ implied by Eq.~(\ref{e}) holds in the Newtonian adiabatic limit with negligible gravitational-wave emission, in which $J_r$ and $J_\theta$ remain conserved. In the numerical integrations of Section~\ref{sect4} we instead include the effects of gravitational radiation on the orbital motion and take nonzero initial eccentricity, namely $e_0=0.1$ for exponential mass loss, and $e_0=0.2$ for linear and Lander models, respectively. The observed decrease of $e$ during inspiral is therefore due to gravitational wave driven circularization, and the adiabatic invariants are not imposed in the numerical implementation.

The time variation of the orbital period of the binary system can be obtained from  Kepler's third law as \cite{Cheng1}
\bea
\frac{1}{P_b(t)}\frac{dP_b(t)}{dt}&=&\frac{3}{2a(t)}-\frac{1}{2M(t)}\frac{dM(t)}{dt}=-\frac{3}{2}\frac{1}{E}\frac{dE}{dt}\nonumber\\
&&+\frac{3}{2}\frac{d}{dt}\ln\left[M_1(t)M_2(t)\right]-\frac{1}{2M(t)}\frac{dM(t)}{dt}.\nonumber\\
\eea

The variation of the orbital period of the binary system can be related to two distinct effects determined by the mass variation of the component objects. The term 
$\frac{3}{2}d\ln\left[M_1(t)M_2(t)\right]/dt-(1/2M(t))(dM(t)/dt)$ follows from Kepler’s third law. We include and keep it in the present general relativistic treatment
of the models with mass variation and energy and angular momentum losses because in the approach we used the masses of the stars in the binary system must represent
the Schwarzschild (or gravitational) masses of the stars. The role of this term in the decay of the orbital period is thus purely Newtonian, and this effect can be described within a classical mechanical description.
However, the main contribution to the period and orbital elements time variation in a binary system is given by the time variation of the quadrupole tensor, due to its  dependence on the time varying  reduced mass $\mu = \mu(t)$. Therefore, the new terms appearing in the expressions of the orbital parameter's decays are essentially  general relativistic in their nature,  and they have their origin in the time derivatives of the time dependent reduced mass, and of the total mass $M(t)$ of the two bodies. The new terms appearing in our formalism are all general relativistic in their nature, and they come from the time derivatives of the sum of masses of the two bodies and of the reduced mass.

Binary pulsars such as PSR B1913+16 \cite{Weis} and PSR B1534+12 \cite{Fonseca} are 
the best sites for the observation of general relativistic effect due to time variation of the mass. The Square Kilometre Array (SKA) telescope, with its high sensitivity in the Southern Hemisphere, will significantly improve the timing precision of recycled pulsars, allowing for a deeper search of testing or detecting potential deviations from general relativity in currently known systems. In \cite{Weis} the variation of the period of the pulsar due to general relativistic effects was determined to be $\dot{P}^{GR}_ b = \left(-2.40263±0.00005\right)\times 10^{-12}$, a result obtained by using the Peters and Mathews results \cite{Peters, Peters1, Hansen}.  On the other hand, the period variation in the time varying masses generalization in the Peters-Mathews formalism gives for the same quantity numerical values of the order of $\dot{P}\approx 10^{-11}$ (exponential mass decay model, see Fig.~\ref{ELPplots_exp}) and $\dot{P}\approx 10^{-10}$, for the Lander-Jones model (see Fig.~\ref{ELPplots_lander}). Hence the present-day observational capabilities could prove, as a result of a systematic search, the possible effects of the mass variation on the gravitational wave emission.

\begin{figure}[htbp!]
	\centering
	\includegraphics[width=0.5\textwidth]{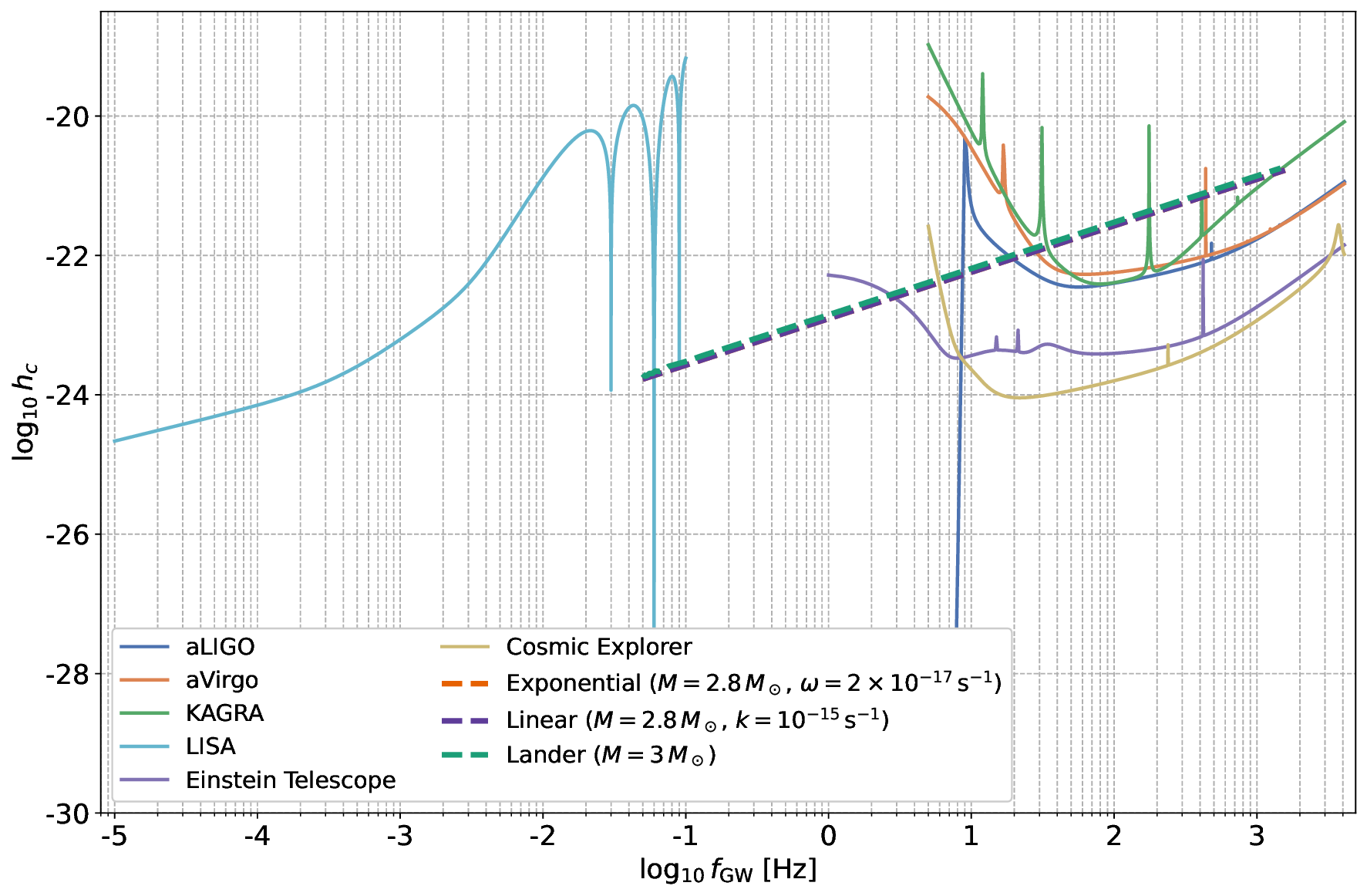}
	\caption{Strain-frequency plot comparing mass-decay models to current and future gravitational-wave detectors.}
	\label{strain_freq}
\end{figure}

The implications of our results for direct gravitational-wave detection can be understood by adopting a reference luminosity distance of $d_0=40$ Mpc. This choice is comparable to the distance of the binary neutron star merger GW170817 \cite{Abbott2, Cantiello2018, Chen2020} and is typical of nearby binary neutron star events targeted by ground-based detectors. The modeled systems fall within the ground-based frequency band, from roughly a few tens of hertz up to about $1.5\,\mathrm{kHz}$, with predicted strain amplitudes of order $|h|\sim 10^{-21}$. They are therefore targets for aLIGO, aVirgo, and KAGRA, as well as next-generation instruments such as the Einstein Telescope and Cosmic Explorer, and may be marginally detectable by space-based missions such as LISA only at the upper end of its frequency range. Pulsar timing arrays such as NANOGrav probe nanohertz frequencies and therefore lie far outside the band considered here. Fig.~\ref{strain_freq} illustrates the characteristic strain of the mass-decay models compared to the detector sensitivity curves at the reference distance $d_0$. The strain scales as $h_c\propto 1/d$, so an approximate maximum detection horizon $d_H$ may be estimated from the peak strain-to-noise ratio. By defining the peak ratio as $\rho_{\max} = \max_f\!\left(h_{c}/h_{n}\right)$, the maximum distance scales as 
$d_H = d_0 \left( \rho_{\max}/\rho_{\mathrm{th}} \right).$
For current-generation detectors at design sensitivity, the peak ratio $\rho_{\max}$ at $40$ Mpc is of order $7-10$. Taking an optimistic threshold $\rho_{\mathrm{th}}=1$ gives a reach of $280-390$ Mpc, whereas a more conservative threshold $\rho_{\mathrm{th}}\simeq 8$ reduces this horizon to $35-49$ Mpc. Next-generation observatories improve the strain sensitivity by a further factor of order $15-20$, pushing the horizon to about $500-1000$ Mpc. The time varying mass models follow nearly identical increasing trends characteristic for the inspiral phase, entering the sensitivity region of the ground-based detectors within the same overall frequency band.

The variation of the orbital period of a binary system depends on the masses of the components,  and on the form of the variation of the mass in time. These two physical quantities depend on the  equation of state of the dense matter of which the stars in binary system are formed. Therefore,  the observations of the effects induced by the change of the gravitational mass in time, and of its impact on the decay via gravitational radiation of the orbital parameters of a binary system could also lead to obtaining some constraints on the equation of state of the dense matter. Moreover, the observations of the time decay of the orbital parameters in binary system with time dependent masses could lead to the possibility of testing other specific general relativistic effects, and bring some novel insights in the physical properties of compact objects.

\section*{Acknowledgements} 

We would like to thank the two anonymous reviewers for comments and suggestions that helped us to significantly improve our manuscript. 

\appendix

\section{Angular-momentum loss from the quadrupole formula}\label{AppLloss}

For completeness we recall a standard phenomenological derivation of the angular-momentum loss associated with the Einstein quadrupole formula \cite{Landau}. Interpreting the averaged energy loss as the work of friction forces $\mathbf{f}$ acting on the source particles,
\begin{gather}
\frac{\D E}{\D t}=\Sa{\bf f}\cdot{\bf v}=\Sa f_i v_i, \label{dE}\\
\frac{\D L^i}{\D t}=\Sa\lp{\bf f}\times{\bf v}\rp^i=\Sa\en^{ijk}x_j f_k, \label{dL}
\end{gather}
and using
\begin{gather}
D_{ij}=\Sa m\lp3x_i x_j-r^2\da_{ij}\rp,\\
\dot{D}_{ij}=\Sa m\lp3v_i x_j+3x_i v_j-2x_k v_k\da_{ij}\rp, \label{dD}
\end{gather}
together with the identity $\langle\dddot D_{ij}\dddot D_{ij}\rangle = \bigl\langle\dot D_{ij}D^{(\rm{v})}_{ij}\bigr\rangle$ for periodic motion, one finds
\begin{equation}
	\frac{dE}{dt} = - \frac{G}{45c^5} \langle \dddot{D}_{ij}\dddot{D}_{ij} \rangle = - \frac{G}{45c^5} \langle \dot{D}_{ij} D^{(\rm{v})}_{ij} \rangle,
\end{equation}
and hence
\be
f_i=-\frac{2G}{15c^5}{D}^{(\rm{v})}_{ij}m x_j.
\ee
Substitution into Eq.~(\ref{dL}) then gives Eq.~(\ref{Lloss}) of the main text.

\section{Orbital decay for constant-mass binaries}\label{AppBinaries}

In this Appendix we record the standard intermediate steps connecting the instantaneous fluxes of Section~\ref{binaries} to the orbit-averaged Peters--Mathews formulae used throughout the paper \cite{Peters, Peters1, Hansen}.

\paragraph{Orbit averages.} With the Keplerian relation $\dot{\vi}=r^{-2}\sqrt{GMa(1-e^2)}$ and period $P_b$ from Eq.~(\ref{period}), the period average of any orbital quantity $Q$ is
\begin{align}
\left\langle Q\right\rangle=\frac{1}{P_b}\int^{P_b}_0 Q\,\D t=\frac{1}{P_b}\int^{2\pi}_0 Q\,\frac{1}{\dot{\vi}}\,\D\vi.
\end{align}
Applying this to Eqs.~(\ref{Eloss}) and (\ref{angloss}) yields
\begin{align}
-\left\langle\frac{\D E}{\D t}\right\rangle
=&\frac{32G^4 M_1^2 M_2^2 M}{5c^5a^5\lp1-e^2\rp^{7/2}}\lp 1+\frac{73}{24}e^2+\frac{37}{96}e^4\rp,
\end{align}
\begin{align}
-\left\langle\frac{\D L}{\D t}\right\rangle
=\frac{32G^3M_1^2M_2^2}{5c^5a^3\lp1-e^2\rp^2}\sqrt{\frac{GM}{a}}\lp1+\frac{7}{8}e^2\rp.
\end{align}

\paragraph{Semi-major axis.} From Eq.~(\ref{avgE}),
\be
\left\langle\frac{\D E}{\D t}\right\rangle=\frac{GM_1M_2}{2a^2}\left\langle\frac{\D a}{\D t}\right\rangle,
\ee
and therefore
\begin{align}
\left\langle\frac{\D a}{\D t}\right\rangle=&\frac{2a^2}{GM_1M_2}\left\langle\frac{\D E}{\D t}\right\rangle\nn\\
=&-\frac{64G^3M_1M_2M}{5c^5a^3\lp1-e^2\rp^{7/2}}\lp1+\frac{73}{24}e^2+\frac{37}{96}e^4\rp, \label{avg_a_app}
\end{align}
which is Eq.~(\ref{avg_a}) of the main text.

\paragraph{Eccentricity and period.} Differentiating $e^2=1+2EL^2M/(G^2M_1^3M_2^3)$ and using the averaged energy and angular-momentum losses gives Eq.~(\ref{avg_ecc}). Finally, Kepler's third law $P_b=2\pi a^{3/2}/\sqrt{GM}$ implies
\begin{align}
\lag\frac{\D P_b}{\D t}\rag=&\frac{3\pi a^{1/2}}{\sqrt{GM}}\lag\frac{\D a}{\D t}\rag,
\end{align}
and hence the expressions for $\langle\dot{P}_b\rangle$ and $\langle\dot{P}_b/P_b\rangle$ quoted in Section~\ref{binaries}.

\section{Quadrupole moments for time varying masses}\label{AppQuad}

Starting from $D_{ij}=D_{cij}f_\mu$ one obtains
\bea
{D}_{ij}&=&{D}_{cij}f_\mu,
\dot{D}_{ij}=\dot{D}_{cij}f_\mu+D_{cij}\dot{f}_\mu,\\
\ddot{D}_{ij}&=&\ddot{D}_{cij}f_\mu+2\dot{D}_{cij}\dot{f}_\mu+D_{cij}\ddot{f}_\mu,\\
\dddot{D}_{ij}&=&\dddot{D}_{cij}f_\mu+3\ddot{D}_{cij}\dot{f}_\mu+3\dot{D}_{cij}\ddot{f}_\mu+D_{cij}\dddot{f}_\mu.
\eea
With $\dot{\vi}=\dot{\vi}_c\sqrt{f_M}$ the constant-mass orbital derivatives transform as
\bea
\dot{D}_{cij}&=&\sqrt{f_M}\lp\dot{D}_{ij}\rp_c,\\
\ddot{D}_{cij}&=&f_M\lp\ddot{D}_{ij}\rp_c+\frac{\dot{f}_M}{2\sqrt{f_M}}\lp\dot{D}_{ij}\rp_c,\\
\dddot{D}_{cij}&=&f_M^{3/2}\lp\dddot{D}_{ij}\rp_c+\frac{3\dot{f}_M}{2}\lp\ddot{D}_{ij}\rp_c
+\frac{2\ddot{f}_Mf_M-\dot{f}_M^2}{4f_M^{3/2}}\lp\dot{D}_{ij}\rp_c\nonumber\\
\eea
which is the matrix ${\bf A}$ of the main text. Writing $\dddot{D}_{ij}={\bf F}_\mu{\bf A}{\bf D}'_c={\bf F}{\bf D}'_c$ and $\ddot{D}_{ij}={\bf F}'_\mu{\bf A}{\bf D}'_c={\bf F}'{\bf D}'_c$ with the row vectors ${\bf F}_\mu=(\dddot{f}_\mu,3\ddot{f}_\mu,3\dot{f}_\mu,f_\mu)$ and ${\bf F}'_\mu=(\ddot{f}_\mu,2\dot{f}_\mu,f_\mu,0)$ yields the coefficients $F_i$ and $F'_i$ quoted in Section~\ref{variable_mass}. Expanding $\dddot{D}_{ij}^2$ then gives
\begin{align}
\dddot{D}_{ij}^2=&(\dddot{D}_{ij})_c^2F_4^2+6(\dddot{D}_{ij})_c(\ddot{D}_{ij})_cF_3 F_4\nn\\
&+6(\dddot{D}_{ij})_c(\dot{D}_{ij})_cF_2F_4+9(\ddot{D}_{ij})_c^2F_3^2\nn\\
&+2(\dddot{D}_{ij})_c(D_{ij})_cF_1 F_4+18(\ddot{D}_{ij})_c(\dot{D}_{ij})_cF_2F_3\nn\\
&+6(\ddot{D}_{ij})_c(D_{ij})_cF_1F_3+9(\dot{D}_{ij})_c^2F_2^2\nn\\
&+6(\dot{D}_{ij})_c(D_{ij})_cF_1F_2+(D_{ij})_c^2F_1^2.
\end{align}

For evaluating the terms containing the products of the time derivatives of the quadrupole moments we obtain
\begin{align}
	&\ddot{D}_{xy}\dddot{D}_{yy}=\lb(\ddot{D}_{xy})_cF'_3+2(\dot{D}_{xy})_cF'_2+(D_{xy})_cF'_1\rb\nn\\
	&\lb(\dddot{D}_{yy})_cF_4+3(\ddot{D}_{yy})_cF_3+3(\dot{D}_{yy})_cF_2+(D_{yy})_cF_1\rb,\nn\\
	&\ddot{D}_{yy}\dddot{D}_{xy}=\lb(\ddot{D}_{yy})_cF'_3+2(\dot{D}_{yy})_cF'_2+(D_{yy})_cF'_1\rb\nn\\
	&\lb(\dddot{D}_{xy})_cF_4+3(\ddot{D}_{xy})_cF_3+3(\dot{D}_{xy})_cF_2+(D_{xy})_cF_1\rb,\nn\\
\end{align}
and
\begin{align}
	&\ddot{D}_{xy}\dddot{D}_{yy}-\ddot{D}_{yy}\dddot{D}_{xy}\nn\\
	=&F_4(\dddot{D}_{yy})_c\lb(\ddot{D}_{xy})_cF'_3+2(\dot{D}_{xy})_cF'_2+(D_{xy})_cF'_1\rb\nn\\
	&-F_4(\dddot{D}_{xy})_c\lb(\ddot{D}_{yy})_cF'_3+2(\dot{D}_{yy})_cF'_2+(D_{yy})_cF'_1\rb\nn\\
	&+3(\ddot{D}_{xy})_c(\dot{D}_{yy})_c(F'_3F_2-2F'_2F_3)\nn\\
	&-3(\ddot{D}_{yy})_c(\dot{D}_{xy})_c(F'_3F_2-2F'_2F_3)\nn\\
	&+(\ddot{D}_{xy})_c(D_{yy})_c(F'_3F_1-3F'_1F_3)\nn\\
	&-(\ddot{D}_{yy})_c(D_{xy})_c(F'_3F_1-3F'_1F_3)\nn\\
	&+(\dot{D}_{xy})_c(D_{yy})_c(2F'_2F_1-3F'_1F_2)\nn\\
	&-(\dot{D}_{yy})_c(D_{xy})_c(2F'_2F_1-3F'_1F_2)\nn\\
	%\end{align}
	%\begin{align}
	=&\lb(\ddot{D}_{xy})_c(\dddot{D}_{yy})_c-(\dddot{D}_{xy})_c(\ddot{D}_{yy})_c\rb F'_3F_4\nn\\
	&+2\lb(\dot{D}_{xy})_c(\dddot{D}_{yy})_c-(\dddot{D}_{xy})_c(\dot{D}_{yy})_c\rb F'_2F_4\nn\\
	&+\lb(D_{xy})_c(\dddot{D}_{yy})_c-(\dddot{D}_{xy})_c(D_{yy})_c\rb F'_1F_4\nn\\
	&+3\lb(\ddot{D}_{xy})_c(\dot{D}_{yy})_c-(\ddot{D}_{yy})_c(\dot{D}_{xy})_c\rb(F'_3F_2-2F'_2F_3)\nn\\
	&+\lb(\dot{D}_{xy})_c(D_{yy})_c-(\dot{D}_{yy})_c(D_{xy})_c\rb(2F'_2F_1-3F'_1F_2)\nn\\
	&+\lb(\ddot{D}_{xy})_c(D_{yy})_c-(\ddot{D}_{yy})_c(D_{xy})_c\rb(F'_3F_1-3F'_1F_3),
\end{align}
respectively. By replacing the indices $(yy)$ with $(xx)$, we obtain $\ddot{D}_{xy}\dddot{D}_{xx}-\ddot{D}_{xx}\dddot{D}_{xy}$.

\section{Coefficients $A_i$, $i=1,2,...,9$}\label{DefA}

The coefficients $A_i$, determining the rate of gravitational wave emission for a binary system with variable mass are defined as follows
\bea
A_1&=&(\dddot{D}_{xx})_c(\ddot{D}_{xx})_c+2(\dddot{D}_{xy})_c(\ddot{D}_{xy})_c\nonumber\\
&&+(\dddot{D}_{yy})_c(\ddot{D}_{yy})_c+(\dddot{D}_{zz})_c(\ddot{D}_{zz})_c,\\
%=&-\frac{6e\mu_c^2}{a\left(1-e^2\right)}\lp\frac{GM_c}{a\lp1-e^2\rp}\rp^{5/2}\sin\vi(1+e\cos\vi)^2\nn\\
%&\left(9e^2\cos{2\vi}+13e^2+40e\cos\vi+18\right),
%\end{align}
%\begin{align}
A_2&=&(\dddot{D}_{xx})_c(\dot{D}_{xx})_c+2(\dddot{D}_{xy})_c(\dot{D}_{xy})_c\nn\\
&&+(\dddot{D}_{yy})_c(\dot{D}_{yy})_c+(\dddot{D}_{zz})_c(\dot{D}_{zz})_c,\\
%=&-\frac{12G^2M_c^2\mu_c^2}{a^2\left(1-e^2\right)^2}\left(e^3\cos{3\vi}+8e^2\cos{2\vi}\rd\nn\\
%&\ld+\left(5e^2+18\right)e\cos{\vi}+10e^2+6\right),
%\end{align}
%\begin{align}
A_3&=&(\ddot{D}_{xx})_c^2+2(\ddot{D}_{xy})_c^2+(\ddot{D}_{yy})_c^2+(\ddot{D}_{zz})_c^2,\\
%=&\frac{6G^2M_c^2\mu_c^2}{a^2\left(1-e^2\right)^2}\left(4e^4+3e^3\cos{3\vi}+20e^2\cos{2\vi}\rd\nn\\
%&\ld+\left(17e^2+36\right)e\cos\vi+20e^2+12\right),
%\end{align}
%\begin{align}
A_4&=&(\dddot{D}_{xx})_c(D_{xx})_c+2(\dddot{D}_{xy})_c(D_{xy})_c\nn\\
&&+(\dddot{D}_{yy})_c(D_{yy})_c+(\dddot{D}_{zz})_c(D_{zz})_c,\\
%=&-12eGM_c\mu_c^2\sqrt{\frac{GM_c}{a(1-e^2)}}\sin\vi,
%\end{align}
%\begin{align}
A_5&=&(\ddot{D}_{xx})_c(\dot{D}_{xx})_c+2(\ddot{D}_{xy})_c(\dot{D}_{xy})_c\nn\\
&&+(\ddot{D}_{yy})_c(\dot{D}_{yy})_c+(\ddot{D}_{zz})_c(\dot{D}_{zz})_c,\\
%=&24e^2GM_c\mu_c^2\sqrt{\frac{GM_c}{a(1-e^2)}}\frac{\sin\vi(e+\cos\vi)}{1+e\cos\vi},
%\end{align}
%\begin{align}
A_6&=&(\ddot{D}_{xx})_c(D_{xx})_c+2(\ddot{D}_{xy})_c(D_{xy})_c\nn\\
&&+(\ddot{D}_{yy})_c(D_{yy})_c+(\ddot{D}_{zz})_c(D_{zz})_c,\\
%=&-\frac{3a\left(1-e^2\right)GM_c\mu_c^2}{(1+e\cos\vi)^2}\nn\\
%&\left(3e^2\cos{2\vi}-e^2+8e\cos\vi+6\right),
%\end{align}
%\begin{align}
A_7&=&(\dot{D}_{xx})_c^2+2(\dot{D}_{xy})_c^2+(\dot{D}_{yy})_c^2+(\dot{D}_{zz})_c^2,\\
%=&\frac{3a\left(1-e^2\right)GM_c\mu_c^2}{(1+e\cos\vi)^2}\times \nn\\
%&\left(-e^2\cos{2\vi}+7e^2+12e\cos\vi+6\right),
%\end{align}
%\begin{align}
A_8&=&(\dot{D}_{xx})_c(D_{xx})_c+2(\dot{D}_{xy})_c(D_{xy})_c\nonumber\\
&&+(\dot{D}_{yy})_c(D_{yy})+(\dot{D}_{zz})_c(D_{zz}),\\
%=&12ea^2(1-e^2)^2\mu_c^2\sqrt{a(1-e^2)GM_c}\frac{\sin\vi}{(1+e\cos\vi)^{3}},
%\end{align}
%and
%\begin{align}
A_9&=&(D_{xx})_c^2+2(D_{xy})_c^2+(D_{yy})_c^2+(D_{zz})_c^2. 
%=&\frac{6a^4(1-e^2)^4\mu_c^2}{(1+e\cos\vi)^4},
\eea

By taking into account the expressions (\ref{Dxx})
%\bea
%&&D_{xx}=\mu r^2(3\cos^2\vi-1), \quad D_{yy}=\mu r^2(3\sin^2\vi-1) \nn \\
%&&D_{xy}=\mu r^2(3\sin\vi\cos\vi), \quad D_{zz}=-\mu r^2,
%\eea
the coefficients $A_1-A_9$ are computed as follows
\begin{align}
A_1=&(\dddot{D}_{xx})_c(\ddot{D}_{xx})_c+2(\dddot{D}_{xy})_c(\ddot{D}_{xy})_c\nn\\
&+(\dddot{D}_{yy})_c(\ddot{D}_{yy})_c+(\dddot{D}_{zz})_c(\ddot{D}_{zz})_c\nn\\
=&-\frac{6e\mu_c^2}{a\left(1-e^2\right)}\lp\frac{GM_c}{a\lp1-e^2\rp}\rp^{5/2}\sin\vi(1+e\cos\vi)^2\nn\\
&\left(9e^2\cos{2\vi}+13e^2+40e\cos\vi+18\right),
\end{align}
\begin{align}
A_2=&(\dddot{D}_{xx})_c(\dot{D}_{xx})_c+2(\dddot{D}_{xy})_c(\dot{D}_{xy})_c\nn\\
&+(\dddot{D}_{yy})_c(\dot{D}_{yy})_c+(\dddot{D}_{zz})_c(\dot{D}_{zz})_c\nn\\
=&-\frac{12G^2M_c^2\mu_c^2}{a^2\left(1-e^2\right)^2}\left(e^3\cos{3\vi}+8e^2\cos{2\vi}\rd\nn\\
&\ld+\left(5e^2+18\right)e\cos{\vi}+10e^2+6\right),
\end{align}
\begin{align}
A_3=&(\ddot{D}_{xx})_c^2+2(\ddot{D}_{xy})_c^2+(\ddot{D}_{yy})_c^2+(\ddot{D}_{zz})_c^2\nn\\
=&\frac{6G^2M_c^2\mu_c^2}{a^2\left(1-e^2\right)^2}\left(4e^4+3e^3\cos{3\vi}+20e^2\cos{2\vi}\rd\nn\\
&\ld+\left(17e^2+36\right)e\cos\vi+20e^2+12\right),
\end{align}
\begin{align}
A_4=&(\dddot{D}_{xx})_c(D_{xx})_c+2(\dddot{D}_{xy})_c(D_{xy})_c\nn\\
&+(\dddot{D}_{yy})_c(D_{yy})_c+(\dddot{D}_{zz})_c(D_{zz})_c\nn\\
=&-12eGM_c\mu_c^2\sqrt{\frac{GM_c}{a(1-e^2)}}\sin\vi,
\end{align}
\begin{align}
A_5=&(\ddot{D}_{xx})_c(\dot{D}_{xx})_c+2(\ddot{D}_{xy})_c(\dot{D}_{xy})_c\nn\\
&+(\ddot{D}_{yy})_c(\dot{D}_{yy})_c+(\ddot{D}_{zz})_c(\dot{D}_{zz})_c\nn\\
=&24e^2GM_c\mu_c^2\sqrt{\frac{GM_c}{a(1-e^2)}}\frac{\sin\vi(e+\cos\vi)}{1+e\cos\vi},
\end{align}
\begin{align}
A_6=&(\ddot{D}_{xx})_c(D_{xx})_c+2(\ddot{D}_{xy})_c(D_{xy})_c\nn\\
&+(\ddot{D}_{yy})_c(D_{yy})_c+(\ddot{D}_{zz})_c(D_{zz})_c\nn\\
=&-\frac{3a\left(1-e^2\right)GM_c\mu_c^2}{(1+e\cos\vi)^2}\nn\\
&\left(3e^2\cos{2\vi}-e^2+8e\cos\vi+6\right),
\end{align}
\begin{align}
A_7=&(\dot{D}_{xx})_c^2+2(\dot{D}_{xy})_c^2+(\dot{D}_{yy})_c^2+(\dot{D}_{zz})_c^2\nn\\
=&\frac{3a\left(1-e^2\right)GM_c\mu_c^2}{(1+e\cos\vi)^2}\times \nn\\
&\left(-e^2\cos{2\vi}+7e^2+12e\cos\vi+6\right),
\end{align}
\begin{align}
A_8=&(\dot{D}_{xx})_c(D_{xx})_c+2(\dot{D}_{xy})_c(D_{xy})_c\nn\\
&+(\dot{D}_{yy})_c(D_{yy})+(\dot{D}_{zz})_c(D_{zz})\nn\\
=&12ea^2(1-e^2)^2\mu_c^2\sqrt{a(1-e^2)GM_c}\frac{\sin\vi}{(1+e\cos\vi)^{3}},
\end{align}
and
\begin{align}
A_9=&(D_{xx})_c^2+2(D_{xy})_c^2+(D_{yy})_c^2+(D_{zz})_c^2\nn\\
=&\frac{6a^4(1-e^2)^4\mu_c^2}{(1+e\cos\vi)^4},
\end{align}
respectively.

\section{Calculation of the integrals in the average gravitational energy loss rate}\label{app1}

 The integrals appearing in Eq.~(\ref{eqloss}) are given by

\begin{align}
&\int^{2\pi}_0\frac{A_1\D \vi}{(1+e\cos\vi)^2}=0,\\
&\int^{2\pi}_0\frac{A_2\D \vi}{(1+e\cos\vi)^2}=-\frac{48\pi G^2M_c^2\mu_c^2\lp4-\sqrt{1-e^2}\rp}{a^2(1-e^2)^2},\\
&\int^{2\pi}_0\frac{A_3\D \vi}{(1+e\cos\vi)^2}=\frac{48\pi G^2M_c^2\mu_c^2\lp4-\sqrt{1-e^2}\rp}{a^2(1-e^2)^2},\\
&\int^{2\pi}_0\frac{A_4\D \vi}{(1+e\cos\vi)^2}=0,\\
&\int^{2\pi}_0\frac{A_5\D \vi}{(1+e\cos\vi)^2}=0,\\
&\int^{2\pi}_0\frac{A_6\D \vi}{(1+e\cos\vi)^2}=-\frac{12\pi GM_c\mu_c^2a\lp3-e^2\rp}{(1-e^2)^{3/2}},\\
&\int^{2\pi}_0\frac{A_7\D \vi}{(1+e\cos\vi)^2}=\frac{12\pi GM_c\mu_c^2a\lp3-e^2\rp}{(1-e^2)^{3/2}},\\
&\int^{2\pi}_0\frac{A_8\D \vi}{(1+e\cos\vi)^2}=0,\\
&\int^{2\pi}_0\frac{A_9\D \vi}{(1+e\cos\vi)^2}=\frac{3\pi \mu_c^2a^4\lp8+40e^2+15e^4\rp}{2(1-e^2)^{3/2}}.
\end{align}

\section{Coefficients $B_i$, $i=1,...,5$}\label{DefB}

The coefficients $B_i$, $i=1,...,5$ are defined as
\begin{align}
B_1=&2\lb(\dot{D}_{xy})_c\lp(\dddot{D}_{yy})_c-(\dddot{D}_{xx})\rp\rd\nn\\
&\ld-(\dddot{D}_{xy})_c\lp(\dot{D}_{yy})_c-(\dot{D}_{xx})_c\rp\rb,\\
B_2=&(D_{xy})_c\lp(\dddot{D}_{yy})_c-(\dddot{D}_{xx})_c\rp \nn\\
&-(\dddot{D}_{xy})_c\lp(D_{yy})_c-(D_{xx})_c\rp,\\
B_3=&3\lb(\ddot{D}_{xy})_c\lp(\dot{D}_{yy})_c-(\dot{D}_{xx})_c\rp\rd\nn\\
&\ld-(\dot{D}_{xy})_c\lp(\ddot{D}_{yy})_c-(\ddot{D}_{xx})_c\rp\rb,\\
B_4=&(\dot{D}_{xy})_c\lp(D_{yy})_c-(D_{xx})_c\rp\nn\\
&-(D_{xy})_c\lp(\dot{D}_{yy})_c-(\dot{D}_{xx})_c\rp,\\
B_5=&(\ddot{D}_{xy})_c\lp(D_{yy})_c-(D_{xx})_c\rp\nn\\
&-(D_{xy})_c\lp(\ddot{D}_{yy})_c-(\ddot{D}_{xx})_c\rp.
\end{align}

The expressions of the coefficients $B_i$, $i=1,...,5$ can be obtained as
\begin{align}
B_1=&-\frac{108G^2M_c^2\mu_c^2e}{a^2\lp1-e^2\rp^2}\sin\vi\lp1+e\cos\vi\rp^2,\\
B_2=&-36GM_c\mu_c^2\sqrt{\frac{GM_c}{a(1-e^2)}}\lp1+e\cos\vi\rp,\\
B_3=&-54GM_c\mu_c^2\sqrt{\frac{GM_c}{a(1-e^2)}}\lp2+e^2+3e\cos\vi\rp,\\
B_4=&-\frac{9\mu^2\sqrt{GM_ca(1-e^2)}a^2(1-e^2)^2}{\lp1+e\cos\vi\rp^2},\\
B_5=&-18GM_c\mu_c^2ae(1-e^2)\frac{\sin\vi}{1+e\cos\vi}.
\end{align} 

\section{Calculation of the integrals in the expression of the angular momentum loss}\label{app2}

The integrals appearing in Eq.~(\ref{eqangloss}) are given by
\begin{align}
&\int^{2\pi}_0\frac{B_1\D \vi}{(1+e\cos\vi)^2}=0,\\
&\int^{2\pi}_0\frac{B_2\D \vi}{(1+e\cos\vi)^2}=-\frac{72\pi GM_c\mu_c^2}{1-e^2}\sqrt{\frac{GM_c}{a}},\\
&\int^{2\pi}_0\frac{B_3\D \vi}{(1+e\cos\vi)^2}=-\frac{216\pi GM_c\mu_c^2}{1-e^2}\sqrt{\frac{GM_c}{a}},\\
&\int^{2\pi}_0\frac{B_4\D \vi}{(1+e\cos\vi)^2}=-\frac{9\pi\mu_c^2\sqrt{GM_ca}a^2(2+3e^2)}{1-e^2},\\
&\int^{2\pi}_0\frac{B_5\D \vi}{(1+e\cos\vi)^2}=0.
\end{align}

\section{The averaged eccentricity}\label{ecc}

The average evolution of the eccentricity can be calculated starting from the definition

\begin{align}
	&2e\left\langle\frac{\D e}{\D t}\right\rangle=\frac{2M_c}{G^2M_{c1}^3M_{c2}^3}\lp\frac{f_M}{f_1^3f_2^3}L^2\left\langle\frac{\D E}{\D t}\right\rangle\rd\nn\\
	&\ld+2EL\frac{f_M}{f_1^3f_2^3}\left\langle\frac{\D L}{\D t}\right\rangle+EL^2\frac{\dot{f}_M-3f_M\lp\frac{\dot{f}_1}{f_1}+\frac{\dot{f}_2}{f_2}\rp}{(f_1f_2)^3}\rp\nn\\
	=&\frac{2M_c}{G^2M_{1c}^3M_{2c}^3}\lb\frac{GM_{1c}^2M_{2c}^2a(1-e^2)}{M_c}\frac{1}{f_1f_2}\lp F_4^2\lag\frac{\D E}{\D t}\rag_c\rd\rd\nn\\
	&-\frac{G\mu_c^2}{10c^5}\lp\frac{16G^2M_c^2\lp4-\sqrt{1-e^2}\rp}{a^2\sqrt{1-e^2}}\lp3F_3^2-2F_2F_4\rp\rd\nn\\
	&+4GM_ca\lp3-e^2\rp\lp3F_2^2-2F_1F_3\rp\nn\\
	&\ld\ld+\frac{1}{6}a^4\lp8+40e^2+15e^4\rp F_1^2\rp\rp\nn\\
	&-GM_{1c}^2M_{2c}^2\sqrt{\frac{G(1-e^2)}{M_ca}}\frac{\sqrt{f_M}}{f_1f_2}\nn\\
	&\lp F'_3 F_4\lag\frac{\D L_z}{\D t}\rag_c+\frac{G\mu_c^2}{5c^5}\sqrt{\frac{GM_c(1-e^2)}{a}}\rd\nn\\
	&\lp8GM_c(F'_1F_4+3F'_3F_2-6F'_2F_3)\rd\nn\\
	&\ld\ld+a^3(2+3e^2)\lp2F'_2F_1-3F'_1F_2\rp\rp\rp\nn\\
	&\ld-\frac{G^2M_{1c}^3M_{2c}^3(1-e^2)}{2M_c}\lp\frac{\dot{f}_M}{f_M}-3\frac{\dot{f}_1}{f_1}-3\frac{\dot{f}_2}{f_2}\rp\rb.
\end{align}

Note that $\sqrt{f_M}F'_3=f_M^{3/2}f_\mu=F_4$, so after canceling and simplification, we have
\begin{align}\label{dedt}
	\left\langle\frac{\D e}{\D t}\right\rangle=&\frac{F_4^2}{f_1f_2}\left\langle\frac{\D e}{\D t}\right\rangle_c-\frac{\mu_c a(1-e^2)}{10c^5e}\frac{1}{f_1f_2}\nn\\
	&\lb\frac{16G^2M_c\lp4-\sqrt{1-e^2}\rp}{a^2\sqrt{1-e^2}}\lp3F_3^2-2F_2F_4\rp\rd\nn\\
	&+4Ga\lp3-e^2\rp\lp3F_2^2-2F_1F_3\rp\nn\\
	&\ld+\frac{a^4}{6M_c}\lp8+40e^2+15e^4\rp F_1^2\rb\nn\\
	&-\frac{G\mu_c(1-e^2)}{5c^5ea}\frac{\sqrt{f_M}}{f_1f_2}\nn\\
	&\lb8GM_c(F'_1F_4+3F'_3F_2-6F'_2F_3)\rd\nn\\
	&\ld+a^3(2+3e^2)\lp2F'_2F_1-3F'_1F_2\rp\rb\nn\\
	&-\frac{1-e^2}{2e}\lp\frac{\dot{f}_M}{f_M}-3\frac{\dot{f}_1}{f_1}-3\frac{\dot{f}_2}{f_2}\rp. 
\end{align}

\section{Approximate coalescence equations}\label{AppCoalApprox}

In the linear model $f=1+kt$ with $kt\ll 1$, the post-Newtonian scaling $GM/(c^2a)\sim\varepsilon^2$ reduces Eq.~(\ref{eqlin}) to
\bea
\frac{\dot{a}}{c}&\sim &-\en^6(1+kt)^3+\frac{2}{1+kt}\frac{ka}{c}-
\en^2\frac{k^4a^4}{c^4(1+kt)^3}-\en^4\frac{k^2a^2}{c^2}.\nonumber\\
\eea
With $a/c\sim\varepsilon P_b$ and $P_b\sim\varepsilon^n t$ at early inspiral (small $n$), the first two terms dominate, giving Eq.~(\ref{eqapp1}). An analogous expansion of Eq.~(\ref{eqexp}) for $f=e^{\omega t}$ with $|\omega t|\ll 1$ and $\omega P_b\ll 1$ retains the Newtonian mass-loss contributions and gives the reduced equation used in the main text.

\end{document}